\documentclass[%
 reprint,
 amsmath,amssymb,
 aps,
]{revtex4-2}

\usepackage{graphicx}
\usepackage{hyperref}
\usepackage{physics}
\usepackage[caption=false]{subfig}
\usepackage{booktabs}
\usepackage{dcolumn}
\usepackage{xcolor}

\immediate\write18{texcount -inc -incbib
-sum borra.tex > /tmp/wordcount.tex}

\immediate\write18{texcount -char -freq
 borra.tex > /tmp/charcount.tex}

\begin{document}

\preprint{APS/123-QED}

\title{Diagonalization of Hamiltonian for finite-sized dispersive media: Canonical quantization with numerical mode-decomposition (CQ-NMD)}
\author{Dong-Yeop Na}
\affiliation{%
School of Electrical and Computer Engineering, Purdue University, West Lafayette IN, 47907, USA\\
}%

\author{Jie Zhu}%
\affiliation{%
School of Electrical and Computer Engineering, Purdue University, West Lafayette IN, 47907, USA\\
}%

\author{Weng C. Chew}
\email{wcchew@purdue.edu; the authors are with Purdue Quantum Science Engineering Institute.}
\affiliation{%
School of Electrical and Computer Engineering, Purdue University, West Lafayette IN, 47907, USA\\
}%

\date{\today}

\begin{abstract}
We present a new math-physics modeling approach, called {\it canonical quantization with numerical mode-decomposition}, for capturing the physics of how incoming photons interact with finite-sized dispersive media, which is not describable by the previous Fano-diagonalization methods.
The main procedure is to (1) study a system where electromagnetic (EM) fields are coupled to non-uniformly-distributed Lorentz oscillators in Hamiltonian mechanics, (2) derive a generalized Hermitian eigenvalue problem for conjugate pairs in coordinate space, (3) apply computational electromagnetics methods to find a countably-finite set of time-harmonic eigenmodes that diagonalizes the Hamiltonian, and (4) perform the subsequent canonical quantization with mode-decomposition.
Moreover, we provide several numerical simulations that capture the physics of full quantum effects, impossible by classical Maxwell's equations, such as non-local dispersion cancellation of an entangled photon pair and Hong-Ou-Mandel (HOM) effect in a dispersive beam splitter.

\end{abstract}

\maketitle


\section{Introduction}
\subsection{Main contribution}
\begin{figure}
\centering
\includegraphics[width=\linewidth]
{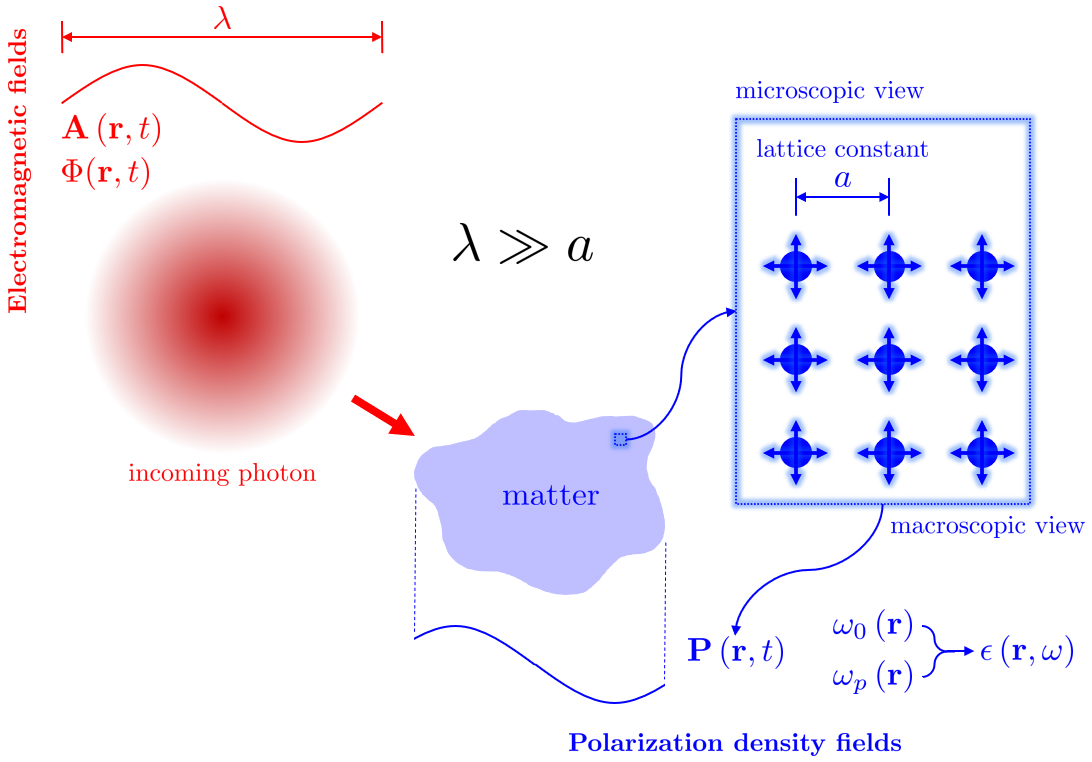}
\caption{2-D illustration of a problem geometry where electromagnetic fields are coupled to lossless Lorentz oscillators forming polarization density fields inside a macroscopic dispersive and inhomogeneous dielectric medium.}
\label{fig:schm_LO}
\end{figure}

We present a new math-physics modeling approach, {\it canonical quantization with numerical mode-decomposition} (CQ-NMD), suited for studying how incoming (entangled) photons interact with finite-sized dispersive media (see Fig. \ref{fig:schm_LO}), which are not simply described by the previous Fano-diagonalization methods.
To do this, we shall (1) study a system where electromagnetic (EM) fields are coupled to non-uniformly-distributed Lorentz oscillators in Hamiltonian mechanics, (2) derive a generalized Hermitian eigenvalue problem (GH-EVP) for conjugate pairs in the coordinate space, (3) apply computational electromagnetics (CEM) methods to find a countably-finite set of time-harmonic eigenmodes which diagonalize the Hamiltonian, and (4) perform the subsequent canonical quantization with mode-decomposition.
We consider two applications of this modeling for fully quantum effects including the Hong-Ou-Mandel (HOM) effects in a dispersive beam splitter, and non-local dispersion cancellation (NLDC) for an energy-time entangled photon pair, showing that such CEM-driven quantum electromagnetics/optics (QEM/QO) research has a great promise.  To our knowledge, this is the first time that non-local dispersion cancellation has been modeled by a numerical method that can be applied to geometry of arbitrary complexity.

Pioneering theoretical works \cite{Knoll1987action,Glauber1991quantum} have shown canonical quantization schemes for dispersionless, lossless, and inhomogeneous dielectric media.
In essence, the underlying principle is the same as that of free fields.
Furthermore, it is shown in our recent study \cite{Na2020quantum} that solving for the eigenmodes can be numerically performed by exploiting CEM methods.
Such CEM-driven QEM/QO simulations have a great potential for effectively dealing with practical QEM/QO applications involving arbitrary geometric complexity, such as in quantum imaging, sensing, and radar.

According to \cite{Jauslin2020Critical}, the free-field contribution should be added to the previous Fano-diagonalization-based quantization scheme so that one can accurately model finite-sized media illuminated by incoming photons from the free space.
The complete description was recently proposed by \cite{Jauslin2019canonical} in momentum (or spectral) space.
We show that our formulation is mathematically equivalent to theirs, though, our formulation is in the coordinate space with the use of CEM methods.
Thus, the proposed approach can tackle arbitrary geometrical complexity present in finite-sized dispersive media.
Also, our work and derivations are based on sound mathematical logic and validated with numerical studies.  Then we use our math-physics model to reproduce the ``weird" physical phenomena that have been reported in the literature.

Our main contributions are three folds:
\begin{itemize}
  \item We derive a generalized Hermitian eigenvalue problem (GH-EVP) for electromagnetic (EM) fields coupled to non-uniformly-distributed Lorentz oscillators, which model finite-sized dispersive media, directly in coordinate space.
  \item We exploit computational electromagnetics (CEM) methods to solve the GH-EVP with arbitrary geometric complexity to obtain a countably-finite set of time-harmonic eigenmodes that diagonalizes the Hamiltonian; hence, the subsequent quantization becomes easier.
  \item Our approach is suitable for studying interaction between incoming (entangled) photons from the free space with arbitrary finite-sized dispersive media, such as quantum plasmonic devices or quantum low-loss optical components. These cannot be modeled by the previous Fano-diagonalization methods.
\end{itemize}

Also, from our model, we can see clearly the dressing of the modes of the system due to coupling between free-field modes and the material modes.  Moreover, we can clearly see from the math that when the material medium is removed or shrunked to zero, we retrieve the free-field modes and vice-versa.  This is not easy to observe when Fano diagonalization approach is used.

Although we present the GH-EVP for the generalized Lorenz gauge, all numerical simulations are performed with the Coulomb gauge.  Thus it is equivalent to the Lorenz gauge with zero scalar potential, to reduce the redundancy of the longitudinal component of the vector potential.

We use the Bloch-periodic boundary conditions (B-PBC) on the GH-EVP to simulate an infinite region problem.  When the period tends to infinity, we retrieve the open infinite region case.  Hence, the GH-EVP is exactly Hermitian.
The B-PBC is the generalized version of conventional PBC, allowing one to extract eigenmodes in the traveling-wave form in the presence of arbitrary, lossless, inhomogeneous media.
As a result, the subsequent quantization procedure becomes mathematically homomorphic to that of the free space.  It is to be noted that despite that we are working with lossless media, the Kramers-Kronig relation is still satisfied \cite{Poon2009}.

%
%

\subsection{Reviews on previous macroscopic quantum electrodynamics works}

The quantum nature of EM fields is basically captured by solving the quantum Maxwell's equations (QME) \cite{Mandel1995optical,cohen1998atom,SCHEEL2008numerical,CHEW2016quantum}, together with solving the quantum state equation.
In the QME, classical Maxwell field and source variables are elevated to infinite-dimensional quantum {\it operators}, expressible by
\begin{flalign}
\nabla \times \hat{\mathbf{E}}(\mathbf{r},t)&=-\frac{\partial \hat{\mathbf{B}}(\mathbf{r},t)}{\partial t},
\nonumber
\\
\nabla \times \hat{\mathbf{H}}(\mathbf{r},t)&=\hat{\mathbf{J}}(\mathbf{r},t)+\frac{\partial \hat{\mathbf{D}}(\mathbf{r},t)}{\partial t},
\nonumber
\\
\nabla \cdot \hat{\mathbf{D}}(\mathbf{r},t)&=\hat{\rho}(\mathbf{r},t),
\nonumber
\\
\nabla \cdot \hat{\mathbf{B}}(\mathbf{r},t)&=0,
\label{eq:QME}
\end{flalign}
The quantum state equations (QSE)
taking the form of
\begin{flalign}
\hat{H}\ket{\psi}
=
i\hbar\frac{\partial}{\partial t}\ket{\psi}
\label{eq:QSE}
\end{flalign}
where $\hat{H}$ and $\ket{\psi}$ denote Hamiltonian operator and quantum state, respectively.
This equation is often called Schrodinger equation, but when the Hamiltonian is replaced by the Dirac Hamiltonian, then it is often called Dirac equation \cite{dirac1981principles}.  To avoid the confusion, we shall call it the quantum state equation, since we are using an electromagnetic Hamiltonian which is very different from Schrodinger's original Hamiltonian.

With the quantum state known, one can evaluate the expectation value or variance of observables.
Note that above QMEs are rigorously derived in the Heisenberg picture, in the coordinate space, for inhomogeneous and anisotropic media when impressed sources are present \cite{CHEW2016quantum}.
The space and time
dependence of field operators obeys QMEs, similar to the classical Maxwellian variables.
On the other hand, the ``weird'' properties such as ``superposition'' and ``entanglement'' can be modeled by solving the QSE.
Having no classical analogue, such properties are the main reason for the ``weird" performance beyond the classical limit.


In place of the atomistic description, the macroscopic theory on quantum electrodynamics (QED), proposed by Jauch and Watson \cite{Jauch1948phenomenological}, is more practical to analyze large-scale quantum technologies.
This framework is valid as long as the wavelength of photons is much larger than a lattice constant \cite{Fano1956atomic,Hopfield1958theory}.
In this, the EM characteristics of a matter (composed of a large number of atoms) are embodied in a phenomenological medium described by the effective permittivity and permeability, as is done in the classical Maxwell's theory.
Thus, it can reduce the significantly
the needed degrees of freedom (DoFs) for modeling.
This approach has been successfully applied to study various quantum-related applications, for instance, quantum metamaterials \cite{PLUMRIDGE2008406}, Casimir forces \cite{Philbin2011Casimir}, spontaneous emission in photonic structures \cite{Pelton2015}, quantum plasmonics \cite{Tame2013}, just to name a few.

Recently, the Jauslin's group has shown that the previous Fano-diagonalization method is incomplete when it comes to studying finite-sized dispersive and dissipative media \cite{Jauslin2020Critical}.
Specifically, in the vanishing limit of finite-sized media, the previous Fano-diagonalization approach, which only includes the medium-assisted field operators, cannot recover the free-field operator due to the absence of the free-field contribution.  Hence, the Fano-diagonalization approach violates a simple sanity check.
The complete quantization formulation was proposed in \cite{Jauslin2019canonical} in momentum space.

But analytic solutions of time-harmonic eigenmodes are often not available.
More importantly, the corresponding Helmholtz wave equation for vector potentials
\begin{flalign}
\nabla\times\frac{1}{\mu_{0}}\nabla\times\tilde{\mathbf{A}}(\mathbf{r})
-
\underbrace{\omega^{2}\epsilon(\mathbf{r},\omega)}_{\text{eigenvalue}}
\tilde{\mathbf{A}}(\mathbf{r})
=
0
\end{flalign}
cannot be converted to a simple explicit eigenvalue problem (EVP) since the eigenvalue $\omega$ is implicit.
As such, the two fundamental properties essential for canonical quantizations of systems does not hold in a strict sense: (1) completeness of eigenmodes and (2) realness of eigenfrequency.
Nevertheless, such implicit EVP may be solved by some {\it ad-hoc} fashions in the past, such as, finite-difference time-domain (FDTD) method, iterative eigenvalue algorithms, or cutting surface method \cite{cuttingsurfacemethod}.  In contrast, we formulate this as an explicit eigenvalue problem here.

\section{Diagonalization of classical Hamiltonian via time-harmonic eigenmodes}
\subsection{Description in Hamiltonian mechanics}
Consider EM fields coupled to a cluster of lossless Lorentz oscillators in the 3-D free space, $V$, as illustrated in Fig. \ref{fig:schm_LO}.
Lorentz (or medium) oscillators can be non-uniformly distributed over $V$, modeling an arbitrary lossless, isotropic, dispersive, and inhomogeneous dielectric medium.

Fundamental dynamical variables are vector and scalar potentials and polarization density field, denoted by $\mathbf{A}$, $\Phi$, and $\mathbf{P}$, respectively.
Suggested in \cite{Sha2018Dissipative}, we define conjugate variables of $\mathbf{A}$, $\Phi$, and $\mathbf{P}$ as
\begin{flalign}
&
\boldsymbol{\Pi}_{AP}\triangleq\epsilon_{0}\frac{\partial \mathbf{A}}{\partial t}-\mathbf{P},~
\Pi_{\Phi}\triangleq\chi_{0}\frac{\partial \Phi}{\partial t},~
\boldsymbol{\Pi}_{P}\triangleq\frac{\beta(\mathbf{r})}{\epsilon_{0}}\frac{\partial \mathbf{P}}{\partial t}.
\label{eqn:CV_pol}
\end{flalign}
The corresponding Hamiltonian is then given by
\begin{flalign}
H
&=
\int_{V}d\mathbf{r}
\mathcal{H}(\mathbf{r},t)
=
\frac{1}{2}
\int_{V}d\mathbf{r}
\Biggl(
\frac{1}{\epsilon_{0}}
\left|
\boldsymbol{\Pi}_{AP}
\right|^{2}
+
\frac{1}{\mu_{0}}
\left|
\nabla\times\mathbf{A}
\right|^{2}
\nonumber \\
&
+
\frac{1}{\chi_0}\left(\nabla\cdot \epsilon_0 \mathbf{A}\right)^{2}
-
\epsilon_{0}\left|\nabla\Phi\right|^{2}
-
\frac{1}{\chi_{0}}\Pi_{\Phi}^{2}
+
\frac{\epsilon_{0}}{\beta(\mathbf{r})}
\left|
\boldsymbol{\Pi}_{P}
\right|^{2}
\nonumber \\
&
+
\frac{f(\mathbf{r})+1}{\epsilon_{0}}
\left|
\mathbf{P}
\right|^{2}
+
\frac{2}{\epsilon_{0}}
\boldsymbol{\Pi}_{AP}\cdot\mathbf{P}
+
2\mathbf{P}\cdot\nabla\Phi
\Biggr)
\label{eqn:original_H}
\end{flalign}
where $f(\mathbf{r})=\omega_{0}^{2}(\mathbf{r})/\omega_{p}^{2}(\mathbf{r})$ and $\beta(\mathbf{r})=1/\omega_{p}^{2}(\mathbf{r})$.
Note that $\omega_{p}(\mathbf{r})$ and $\omega_{0}(\mathbf{r})$ are the plasma and resonant frequencies of a Lorentz oscillator located at $\mathbf{r}$.
Then Hamilton's equations of motion (EoMs) can be explicitly written by \cite{Sha2018Dissipative}
\begin{flalign}
\frac{\partial \mathbf{A}}{\partial t}&
=
\frac{\delta H}{\delta \boldsymbol{\Pi}_{AP}}
=\frac{1}{\epsilon_{0}}\Bigl(\boldsymbol{\Pi}_{AP}+\mathbf{P}\Bigr),
\nonumber
\\
\frac{\partial \boldsymbol{\Pi}_{AP}}{\partial t}&
=
-\frac{\delta H}{\delta \mathbf{A}}
=
-\nabla\times\frac{1}{\mu_{0}}\nabla\times\mathbf{A} +\epsilon_{0}\nabla\frac{1}{\chi_{0}}\nabla\cdot\epsilon_{0}\mathbf{A},
\nonumber
\\
\frac{\partial \Phi}{\partial t}
&
=
-\frac{\delta H}{\delta \Pi_{\Phi}}
=
\frac{1}{\chi_{0}}\Pi_{\Phi},
\nonumber
\\
\frac{\partial \Pi_{\Phi}}{\partial t}
&
=
\frac{\delta H}{\delta \Phi}
=
\nabla\cdot\epsilon_{0}\nabla\Phi-\nabla\cdot\mathbf{P},
\nonumber
\\
\frac{\partial \mathbf{P}}{\partial t}
&=
\frac{\delta H}{\delta \boldsymbol{\Pi}_{P}}
=
\frac{\epsilon_{0}}{\beta(\mathbf{r})}\boldsymbol{\Pi}_{P},
\nonumber
\\
\frac{\partial \boldsymbol{\Pi}_{P}}{\partial t}
&=
-\frac{\delta H}{\delta \mathbf{P}}
=
-\frac{1}{\epsilon_{0}}\boldsymbol{\Pi}_{AP}
-\frac{f(\mathbf{r})+1}{\epsilon_{0}}\mathbf{P}-\nabla\Phi.
\end{flalign}

Defining generalized position and momentum for the whole system as
\begin{flalign}
\mathbf{q}\triangleq \left[\mathbf{A},~\Phi,~\mathbf{P}\right]^{T}, \quad \mathbf{p}\triangleq \left[\boldsymbol{\Pi}_{AP},~ -\Pi_{\Phi},~ \boldsymbol{\Pi}_{P}\right]^{T},
\end{flalign}
one can compactly write the above Hamiltonian in a block matrix form as
\begin{flalign}
H
=
\frac{1}{2}
\int_{V}d\mathbf{r}
\left[
\begin{matrix}
\mathbf{q} \\
\mathbf{p} \\
\end{matrix}
\right]^{\dag}
\cdot
\left[
\begin{matrix}
\overline{\mathbf{K}} & \overline{\mathbf{C}} \\
\overline{\mathbf{C}}^{\dag} & \overline{\mathbf{M}}
\end{matrix}
\right]
\cdot
\left[
\begin{matrix}
\mathbf{q} \\
\mathbf{p} \\
\end{matrix}
\right].
\label{eqn:classical_Ham}
\end{flalign}
where each block matrix can be explicitly written by
\begin{flalign}
\overline{\mathbf{K}}
&=
\left[
\begin{array}{c|c|c}
\nabla\times\frac{1}{\mu_{0}}\nabla\times
-\epsilon_{0}\cdot\nabla\frac{1}{\chi_{0}}\nabla\cdot\epsilon_{0}
&
0
&
0 \\ [.1 cm]\hline \rule{0pt}{.85\normalbaselineskip}
0
&
\nabla\cdot\epsilon_{0}\nabla
&
-\nabla\cdot
\\ \hline \rule{0pt}{1\normalbaselineskip}
0
&
\nabla
&
\frac{f(\mathbf{r})+1}{\epsilon_{0}}
\end{array}
\right],
\nonumber \\
\overline{\mathbf{C}}
&=
\left[
\begin{array}{c|c|c}
0 & 0 & 0
 \\ \hline \rule{0pt}{.85\normalbaselineskip}
0 & 0 & 0
 \\ \hline \rule{0pt}{1\normalbaselineskip}
\frac{1}{\epsilon_{0}} & 0 & 0
\end{array}
\right],
\quad
\overline{\mathbf{M}}
=
\left[
\begin{array}{c|c|c}
\frac{1}{\epsilon_{0}} & 0 & 0
 \\ [.1 cm]\hline \rule{0pt}{1\normalbaselineskip}
0
&
-\frac{1}{\chi_{0}}
&
0
 \\ [.1 cm]\hline \rule{0pt}{1\normalbaselineskip}
0 & 0 & \frac{\beta(\mathbf{r})}{\epsilon_{0}}
\end{array}
\right].
\end{flalign}
Note that in the block matrix representation the partitions are delimited by solid vertical and horizontal lines.
The Hamilton's EoMs can be written in the block matrix form (see its details in Appendix \ref{App_1}) as
\begin{flalign}
\frac{\partial}{\partial t}
\left[
\begin{matrix}
\mathbf{q} \\
\mathbf{p}
\end{matrix}
\right]
=
\left[
\begin{matrix}
\overline{\mathbf{C}}^{\dag} & \overline{\mathbf{M}} \\
-\overline{\mathbf{K}} & -\overline{\mathbf{C}}
\end{matrix}
\right]
\left[
\begin{matrix}
\mathbf{q} \\
\mathbf{p}
\end{matrix}
\right].
\label{eqn:H_EoMs_1}
\end{flalign}
However, the Hamilton's EoMs may not be simply converted to a Hermitian eigenvalue problem due to the presence of $\overline{\mathbf{C}}$ (cross-coupling term), as discussed in Appendix \ref{App_2b} in detail.

\subsection{Generalized Hermitian eigenvalue problem}\label{sec:cross_coupling}
To derive a simpler standard Hermitian eigenvalue problem, we redefine generalized position and momentum such as
\begin{flalign}
\mathbf{q}\triangleq \left[\mathbf{A},~\Pi_{\Phi},~\boldsymbol{\Pi}_{P}\right]^{T},\quad \mathbf{p}\triangleq \left[\boldsymbol{\Pi}_{AP},~\Phi,~-\mathbf{P}\right]^{T},
\end{flalign}
motivated by reducing the cross-coupling term $\overline{\mathbf{C}}$.
It still preserves the structure of the original Hamiltonian density $\mathcal{H}(\mathbf{r},t)$ in \eqref{eqn:original_H}, i.e.,
\begin{flalign}
H
=
\frac{1}{2}
\int_{V}d\mathbf{r}
\left[
\begin{matrix}
\mathbf{q} \\
\mathbf{p}
\end{matrix}
\right]^{\dag}
\cdot
\left[
\begin{matrix}
\overline{\mathbf{K}} & \overline{\mathbf{0}} \\
\overline{\mathbf{0}} & \overline{\mathbf{M}}
\end{matrix}
\right]
\cdot
\left[
\begin{matrix}
\mathbf{q} \\
\mathbf{p}
\end{matrix}
\right]
\label{eqn:d_Hamil}
\end{flalign}
where the ``spring constant'' and ``mass'' matrices are also redefined as
\begin{flalign}
\overline{\mathbf{K}}&=
\left[
\begin{array}{c|c|c}
\nabla\times\frac{1}{\mu_{0}}\nabla\times-\epsilon_{0}\nabla\frac{1}{\chi_{0}}\nabla\cdot\epsilon_{0}
&
0
&
0
 \\ [.1 cm]\hline \rule{0pt}{.85\normalbaselineskip}
0
&
-\frac{1}{\chi_{0}}
&
0
 \\ [.1 cm]\hline \rule{0pt}{1\normalbaselineskip}
0
&
0
&
\frac{\beta(\mathbf{r})}{\epsilon_{0}}
\end{array}
\right],
\\
\overline{\mathbf{M}}&=
\left[
\begin{array}{c|c|c}
\frac{1}{\epsilon_{0}}
&
0
&
\frac{1}{\epsilon_{0}}
 \\ [.1 cm]\hline \rule{0pt}{.85\normalbaselineskip}
0
&
\nabla\cdot\epsilon_{0}\nabla
&
-\nabla\cdot
 \\ [.1 cm]\hline \rule{0pt}{1\normalbaselineskip}
\frac{1}{\epsilon_{0}}
&
\nabla
&
\frac{f(\mathbf{r})+1}{\epsilon_{0}}
\end{array}
\right].
\end{flalign}
As a result, the Hamilton's EoMs can be rewritten as
\begin{flalign}
\frac{\partial}{\partial t}
\left[
\begin{matrix}
\mathbf{q} \\
\mathbf{p}
\end{matrix}
\right]
=
\left[
\begin{matrix}
\overline{\mathbf{0}} & \overline{\mathbf{M}}
\\
-\overline{\mathbf{K}} & \overline{\mathbf{0}}
\end{matrix}
\right]
\cdot
\left[
\begin{matrix}
\mathbf{q} \\
\mathbf{p}
\end{matrix}
\right].
\label{eqn:mc_EoMs}
\end{flalign}
Thus, one can derive EoMs only for the generalized $\mathbf{q}$ or $\mathbf{p}$  variable  involving a second order time derivative.  For instance, for the $\mathbf{q}$ case,
\begin{flalign}
\frac{\partial^{2}}{\partial t^{2}}\mathbf{q}
=
\overline{\mathbf{M}}\cdot\left(-\overline{\mathbf{K}}\right)\cdot\mathbf{q}.
\end{flalign}
Since $\overline{\mathbf{M}}$ and $\overline{\mathbf{K}}$ are both postivie-definite and Hermitian, the above is convertible to an explicit generalized Hermitian eigenvalue problem (GH-EVP) as
\begin{flalign}
\boxed{
\omega^{2}\overline{\mathbf{M}}^{-1}\cdot\tilde{\mathbf{q}}_{\omega,\lambda}(\mathbf{r})
=
\overline{\mathbf{K}}\cdot\tilde{\mathbf{q}}_{\omega,\lambda}(\mathbf{r}).
}
\label{eqn:GHEVP_no_cross_coupling}
\end{flalign}
where $\tilde{\mathbf{q}}_{\omega,\lambda}(\mathbf{r})$ is a time-harmonic eigenmode of $\mathbf{q}(\mathbf{r},t)$ and $\omega$ is eigenfrequency.
Note that the above GH-EVP is equivalent to equation (14) in \cite{Jauslin2019canonical}, as proven in Appendix \ref{App_2c}.
As a result, one can expand the generalized position in terms of time-harmonic eigenmodes
\begin{flalign}
\boxed{
\mathbf{q}(\mathbf{r},t)
=
\int_{\Omega_{+}}d\omega\sum_{\lambda}
\tilde{\mathbf{q}}_{\omega,\lambda}(\mathbf{r})
\underbrace{d_{\omega,\lambda}e^{-i\omega t}}_{d_{\omega,\lambda}(t)}
+
\text{h.c.}
}
\label{eqn:q_no_coupling}
\end{flalign}
where $\Omega_{+}$ denotes the set of positive eigenfrequencies and $\lambda$ denotes degeneracy index, including propagation directions and polarizations.

\subsection{Diagonalization of Hamiltonian}
The GH-EVP inherently possesses the following two orthonormal conditions
\begin{flalign}
\int_{V}d\mathbf{r}
&
\Bigl(
\tilde{\mathbf{q}}_{\omega,\lambda}^{\dag}
\cdot
\overline{\mathbf{M}}^{-1}
\cdot
\tilde{\mathbf{q}}_{\omega',\lambda'}
\Bigr)
=\delta_{\omega,\omega'}\delta_{\lambda,\lambda'},
\\
\int_{V}d\mathbf{r}
&
\Bigl(
\tilde{\mathbf{q}}_{\omega,\lambda}^{\dag}
\cdot
\overline{\mathbf{K}}
\cdot
\tilde{\mathbf{q}}_{\omega',\lambda'}
\Bigr)
=\omega^{2}\delta_{\omega,\omega'}\delta_{\lambda,\lambda'}.
\label{eqn:OT_2_second_app}
\end{flalign}
Substituting \eqref{eqn:q_no_coupling} into \eqref{eqn:d_Hamil} and using the above orthonormal conditions, one can diagonalize the Hamiltonian as
\begin{flalign}
\boxed{
H
=
\frac{1}{2}
\int_{\Omega_{+}}d\omega\sum_{\lambda}
\omega^{2}
\Bigl(
d^{*}_{\omega,\lambda}
d_{\omega,\lambda}
+
d_{\omega,\lambda}
d^{*}_{\omega,\lambda}
\Bigr).
}
\label{eqn:H_diag_M_coupling}
\end{flalign}
The detailed procedure can be found in Appendix \ref{App_2d}.

Note that the present diagonalization procedure shall be called {\it no-cross-coupling description} since we remove the cross-coupling term $\overline{\mathbf{C}}$ by properly redefining generalized position and momentum.
We also present another digonalization strategy, called {\it cross-coupling description}, even in the presence of the cross-coupling term in Appendix \ref{App_2e}.
It turns out that the latter has a twice larger linear system; thus, the no-cross-coupling description here is more computationally efficient.

\section{Quantization by Mode Decomposition}
The subsequent quantization procedure becomes straightforward with the use of time-harmonic eigenmodes \cite{CHEW2016quantum,Na2020quantum}.
Let us elevate the conjugate pairs into operators
\begin{flalign}
{\mathbf{q}}(\mathbf{r},t) \rightarrow \hat{\mathbf{q}}(\mathbf{r},t),
\quad
{\mathbf{p}}(\mathbf{r},t) \rightarrow \hat{\mathbf{p}}(\mathbf{r},t),
\end{flalign}
which satisfy canonical commutator relations:
\begin{flalign}
\left[
\left[\hat{\mathbf{q}}(\mathbf{r},t)\right]_{i},
\left[\hat{\mathbf{p}}(\mathbf{r}',t)\right]_{j}
\right]
&=
i\hbar\delta(\mathbf{r}-\mathbf{r}')\delta_{i,j}\hat{I},
\label{eqn:FCR_1}
\\
\left[
\left[\hat{\mathbf{q}}(\mathbf{r},t)\right]_{i},
\left[\hat{\mathbf{q}}(\mathbf{r}',t)\right]_{j}
\right]
&=
\left[
\left[\hat{\mathbf{p}}(\mathbf{r},t)\right]_{i},
\left[\hat{\mathbf{p}}(\mathbf{r}',t)\right]_{j}
\right]
=
0
\label{eqn:FCR_2}
\end{flalign}
for $i,j\in\left\{1,2,\cdots,7\right\}$ including 3-dimensional vector components of vector potential and polarization density and scalar potential (and their conjugate pairs).
Note that $\left[\hat{\mathbf{q}}(\mathbf{r},t)\right]_{i}$ (or $\left[\hat{\mathbf{p}}(\mathbf{r}',t)\right]_{j}$) stands for $i$-th element of $\mathbf{q}(\mathbf{r},t)$ (or $j$-th element of $\mathbf{p}(\mathbf{r},t)$).
Annihilation and creation operators can be defined by simply elevating modal amplitudes scaled by $\sqrt{\hbar/\omega}$ as
\begin{flalign}
\boxed{
d_{\omega,\lambda}(t)
\rightarrow
\sqrt{\frac{\hbar}{\omega}}
\hat{d}_{\omega,\lambda}(t),
\quad
d^{*}_{\omega,\lambda}(t)
\rightarrow
\sqrt{\frac{\hbar}{\omega}}
\hat{d}^{\dag}_{\omega,\lambda}(t)
}
\end{flalign}
which also satisfies the bosonic commutator relations:
\begin{flalign}
\left[
\hat{d}_{\omega,\lambda}
,
\hat{d}^{\dag}_{\omega',\lambda'}
\right]
&=
\delta_{\omega,\omega'}
\delta_{\lambda,\lambda'}
\hat{I},\\
\left[
\hat{d}_{\omega,\lambda}
,
\hat{d}_{\omega',\lambda'}
\right]
&=
0
=
\left[
\hat{d}^{\dag}_{\omega,\lambda}
,
\hat{d}^{\dag}_{\omega',\lambda'}
\right].
\end{flalign}
Thus, resulting observables are represented by
\begin{flalign}
\boxed{
\hat{\mathbf{q}}(\mathbf{r},t)
=
\int_{\Omega_{+}}d\omega
\sum_{\lambda}
\tilde{\mathbf{q}}_{\omega,\lambda}(\mathbf{r})
\sqrt{\frac{\hbar}{\omega}}
\underbrace{
\hat{d}_{\omega,\lambda}
e^{-i\omega t}
}_{\hat{d}_{\omega,\lambda}(t)}
+\text{h.c.}
}
\label{eqn:obs_2}
\end{flalign}
One can easily check the consistency between canonical commutator relations and bosonic commutator relations by substituting \eqref{eqn:obs_2} into the LHS of \eqref{eqn:FCR_1} and \eqref{eqn:FCR_2}, using the orthonormal properties of time-harmonic eigenmodes, and showing that the resulting LHS of \eqref{eqn:FCR_1} and \eqref{eqn:FCR_2} becomes the RHS of \eqref{eqn:FCR_1} and \eqref{eqn:FCR_2}.

Finally, the Hamiltonian operator, the quantum equivalence of \eqref{eqn:H_diag_M_coupling} that has been elevated to become a quantum operator, can be diagonalized with respect to the ladder operators for both descriptions as
\begin{flalign}
\boxed{
\hat{H}
=
\int_{\Omega_{+}}
d\omega
\sum_{\lambda}
\hbar\omega
\Bigl(
\hat{d}^{\dag}_{\omega,\lambda}
\hat{d}_{\omega,\lambda}
+
\frac{1}{2}\hat{I}
\Bigr)
}
\end{flalign}
where the zero-point energy becomes $E_{0}=\int_{\Omega_+}d\omega\sum_{\lambda}\hbar\omega/2$.
An eigenstate of the corresponding time-independent (stationary) quantum state equation \eqref{eq:QSE} is the multimode-Fock state.  In a word, the Hamiltonian has been decomposed into sum of Hamiltonians of independent harmonic oscillators.  The eigenstate of each individual Hamiltonian is its respective Fock state.  This physical picture is similar to the quantization of electromagnetic field in vacuum using Fourier mode decomposition.  But here, we have used numerically-sought-for modes using CEM rather than Fourier modes.
It should also be noted that this eigenstate does not represent bare eigenstate for neither free EM field nor polarization density but a dressed state which combines the coupling between them.  In a word, the free field modes have been ``dressed" by the matter modes, and vice versa.

\section{Numerical solutions to GH-EVP}
Solving the GH-EVPs in \eqref{eqn:GHEVP_no_cross_coupling} returns the uncountably-infinite set of eigenmodes, which is impossible in practice.
Furthermore, analytic solutions of \eqref{eqn:GHEVP_no_cross_coupling} may not exist in general.
To remedy this, one can use computational electromagnetic (CEM) methods which can be viewed as subspace projection method \cite{Chew2020ece604}.
Here, we refer subspace projection method to a general procedure to approximate an infinite-dimensional solution space $\mathcal{V}$ by a finite-dimensional (countably-finite) one $\mathcal{V}_{d}$ \footnote{Note that the term subspace projection method (also known as Krylov subspace methods) \cite{saad1992numerical} is also used in the numerical linear algebra field, referring to a procedure to deal with large linear systems efficiently.}.
As a consequence, \eqref{eqn:GHEVP_no_cross_coupling}, as is commonly done in numerical linear algebra \cite{golub2013matrix}, becomes a finite-dimensional linear system such as
\begin{flalign}
\boxed{
\overline{\mathbf{M}}^{-1}_{d}
\cdot
\overline{\boldsymbol{\Psi}}_{d}
\cdot
\overline{\boldsymbol{\omega}}^{2}_{d}
=
\overline{\mathbf{K}}_{d}
\cdot
\overline{\boldsymbol{\Psi}}_{d}
}
\label{eqn:disc_GH_evp}
\end{flalign}
where subscript $d$ stands for the approximation by subspace projection method; $\overline{\mathbf{M}}_{d}$ and $\overline{\mathbf{K}}_{d}$ are discrete counterparts of $\overline{\mathbf{M}}$ and $\overline{\mathbf{K}}$, $\overline{\boldsymbol{\omega}}_{d}$ is a diagonal matrix whose elements are eigenfrequencies including degeneracy, and $\overline{\boldsymbol{\Psi}}_{d}$ is a matrix that collects all numerical time-harmonic eigenmodes.
If the dimension of the solution space $\mathcal{V}_{d}$ was $N$, the size of $\overline{\mathbf{M}}_{d}$, $\overline{\mathbf{K}}_{d}$, $\overline{\boldsymbol{\omega}}_{d}$, and $\overline{\boldsymbol{\Psi}}_{d}$ would be $N\times N$.
Most of subspace projection methods employ a mesh on which continuum solutions are sampled by the finite number.
The dimension of the solution space is closely related to the mesh size.
The continuum eigenmode index $(\omega,\lambda)$ is replaced by a single index $n$ which represents $n$-th numerical time-harmonic eigenmodes $\left[\overline{\boldsymbol{\Psi}}_{d}\right]_{:,n}$ having $n$-th eigenfrequency $\left[\overline{\boldsymbol{\omega}}_{d}\right]_{n,n}=\omega_{n}$.
Note that, an element at $i$-th row and $n$-th column of $\overline{\boldsymbol{\Psi}}_{d}$, i.e., $\left[\overline{\boldsymbol{\Psi}}_{d}\right]_{i,n}$, represents $n$-th numerical time-harmonic eigenmode sampled at $i$-th grid point.

\section{Numerical examples}
In this section, we discuss two numerical studies using the proposed quantization scheme: (1) the Hong-Ou-Mandel (HOM) effect \cite{Hong1987measurement} in a 1-D dispersive beam splitter, and (2) non-local dispersion cancellation (NLDC) for an energy-time entangled photon pair \cite{Franson1992nonlocal}.

It should be noted that these are 1-D simulations in which vector potential $\mathbf{A}$ is always transverse (polarized along $z$-axis) to the propagation direction ($x$-axis) while polarization density is also transverse.
Hence, one can use the Lorenz gauge with $\Phi=\Pi_{\Phi}=0$, which is equivalent to the Coulomb gauge.
And, \eqref{eqn:GHEVP_no_cross_coupling} is to be modified accordingly.
As a result, the resulting dynamical variables are $\mathbf{A}(\mathbf{r},t)=\hat{z}A(x,t)$, $\boldsymbol{\Pi}_{AP}(\mathbf{r},t)=\hat{z}\Pi_{AP}(x,t)$, $\mathbf{P}(\mathbf{r},t)=\hat{z}P(x,t)$, and $\boldsymbol{\Pi}_{P}(\mathbf{r},t)=\Pi_{P}(x,t)$.

\subsection{HOM effect in dispersive beam splitter}
Here, we discuss 1-D simulation results of the HOM effect \cite{Hong1987measurement} in a dispersive beam splitter.

\begin{figure*}[ht]
\centering
\includegraphics[width=1\linewidth]
{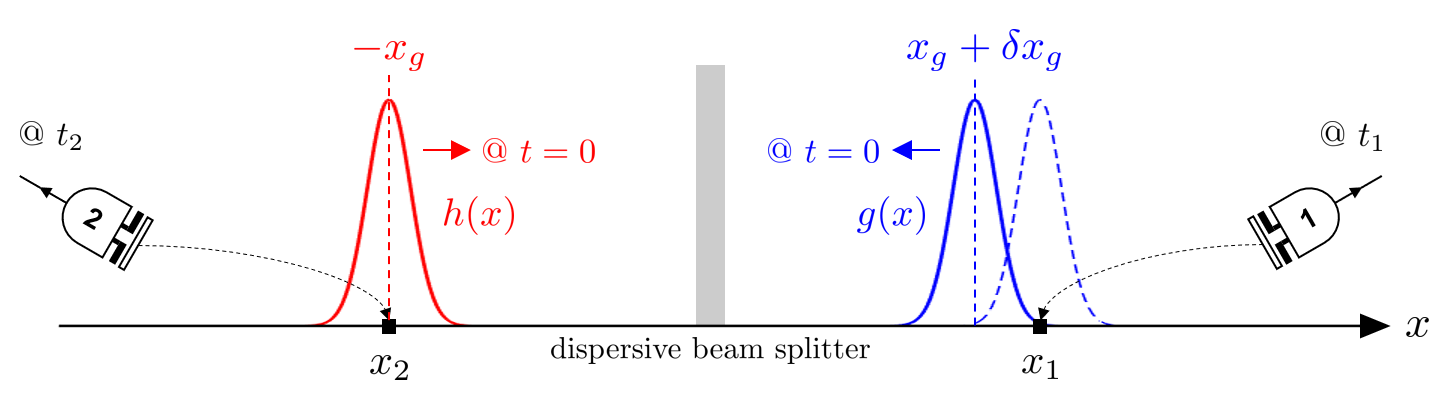}
\caption{
Schematic of 1-D simulations for the HOM effect in a dispersive beam splitter.
A non-entangled photon pair is initially localized around $x_{g}$ and $-x_{g}$ and sent to a beam splitter.
After the interference, we measure the second order correlation at $(x_{1},t_{1})$ and $(x_{2},t_{2})$ by perturbing the initialization position ($\delta x_{g}$) of the photon initialized on the right side.
}
\label{fig:schm_HOM}
\end{figure*}

The problem geometry is illustrated in Fig. \ref{fig:schm_HOM}.
The problem domain $V\in\left\{x\in\left[-L/2,L/2\right]\right\}$, which is the free space, includes the dielectric slab in the middle with thickness $L_s$.
Then, a non-entangled photon pair is initialized ($t=0$) around $x_1=x_{g}$ and $x_2=-x_{g}$ in the free space and sent to a beam splitter.
Each photon is assumed to be polychromatic, riding on a Gaussian wavepacket.
After the interference, we measure the second order correlation at $(x_{1},t_{1})$ and $(x_{2},t_{2})$ by perturbing the initialization position of the photon on the right side, i.e., $x_1=x_{g}+\delta x_{g}$.
The beam splitter is assumed to be made of a single dielectric slab, which is modeled by filling Lorentz oscillators to account for dispersion effects.
Simulation parameters, design of the beam splitter, relevant numerical setup, and modeling incoming polychromatic photons are discussed in detail in Appendix \ref{App_2f}.

\begin{figure}
\centering
\includegraphics[width=\linewidth]
{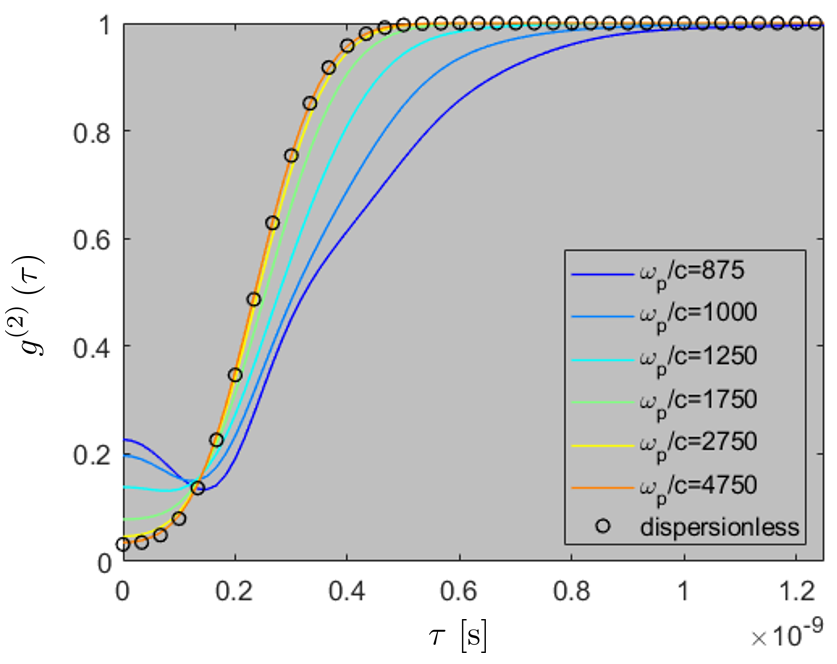}
\caption{
Second order correlation versus time delay ($g^{(2)}(\tau)$) for various $\omega_p$.
When $\tau=0$, dispersion degrades the perfect destructive interference between two incident photons; consequently, $g^{(2)}(\tau)$ increases depending on the dispersion degree.
In contrast, when $\tau \neq 0$, dispersion can mitigate the time-harmonic decoherence such that $g^{(2)}(\tau)$ is slightly lower than the dispersionless case.
}
\label{fig:dispersive_HOM}
\end{figure}
Fig. \ref{fig:dispersive_HOM} shows the second order correlation versus $\tau$ for various $\omega_{p}$.
The smaller $\omega_{p}$, the more dispersive the beam splitter becomes.
Note that the dispersionless case was calculated based on the canonical quantization with numerical mode-decomposition for inhomogeneous and dispersionless media \cite{Na2020quantum}.

The almost zero coincidence when $\tau=0$ is the clear evidence of the creation of path-entangled photons, i.e., $N00N$ state where $N=2$.
This results from the perfect destructive interference between the two photons inside the 50:50 beam splitter \cite{Fearn1987quantum,Prasad1987quantum,Gerry2004introductory}.
When $\tau \neq 0$, the temporal decoherence\textemdash different arrival times of the incident photons to the beam splitter\textemdash degrades the perfect destructive interference; consequently, $g^{(2)}$ gradually increases as $\tau$ gets larger.

In the presence of dispersion, even when $\tau=0$, two polychromatic photons cannot have the perfect destructive interference over their whole bandwidth.
This is because the dispersive beam splitter quickly loses the 50:50 performance as an operating frequency deviates from the carrier frequency of photons.
In other words, the 50:50 performance bandwidth becomes much narrower than the photon's bandwidth depending on the dispersion extent; hence, $g^{(2)}$ increases.
It is interesting to observe that when $\tau\neq 0$, $g^{(2)}$ for dispersive cases gets lower than that of the dispersionless case.
This is because the dispersion effects can mitigate the degradation by the temporal decoherence.
More specifically, the interaction time between photons and the dispersive beam splitter becomes longer than the dispersionless case so that photons can stay longer in the dispersive beam splitter.
As a result, the partial destructive interference can happen even though the arrival times of incidence photons are mismatched.
This is the hallmark of dispersion effects, viz., the decrease of the quality factor of the HOM dip, and our simulation correctly captured this effect.

\begin{figure}
\centering
\subfloat[$\omega_{p}/c=4750$]{\includegraphics[width=1\linewidth]{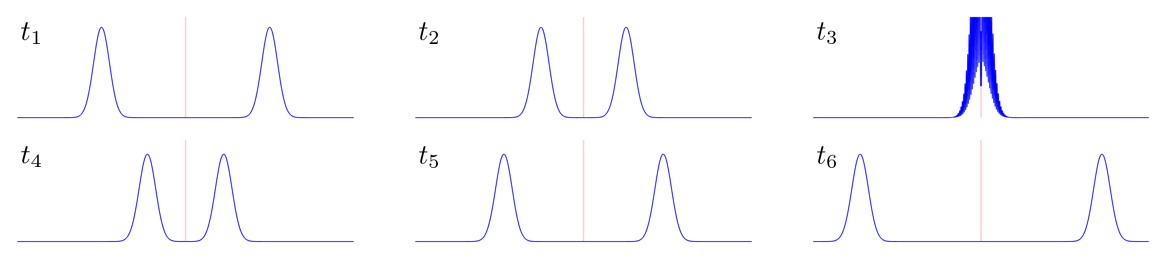}}
\\
\subfloat[$\omega_{p}/c=875$]{\includegraphics[width=1\linewidth]{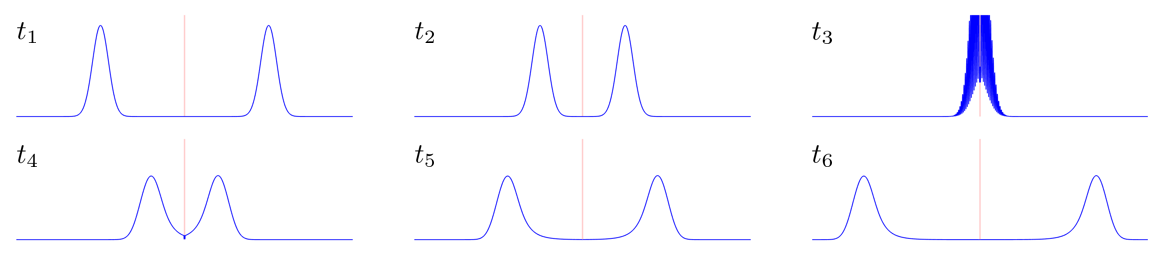}}
\caption{
Time evolution of the energy density expectation value when $\tau=0$ for (a) $\omega_{p}/c=4750$ and (b) $\omega_{p}/c=875$.
The former is almost dispersionless whereas the latter is highly dispersive.
Accordingly, one can observe the deformation of wavepackets in the latter case due to the group velocity dispersion while passing  the beam splitter.
}
\label{fig:energy_density}
\end{figure}

To show this, Fig. \ref{fig:energy_density} compares the time evolution of the energy density expectation value for $\omega_{p}/c=4750$ and $\omega_{p}/c=875$ which correspond to almost dispersionless and highly dispersive cases, respectively.
One can observe the deformation of the wavepackets after passing the dispersive beam splitter, compared with the dispersionless case.
This is because the chromatic dispersion changes the group velocity of the wavepackets during interacting with the dispersive beam splitter.

\subsection{Non-local dispersion cancellation}
Unlike to classical EM pulses, an energy-time entangled photon pair can cancel the dispersion effects in the non-local sense, called {\it non-local dispersion cancellation} (NLDC).
This is another non-classical feature of entangled photons, first proposed by Franson \cite{Franson1992nonlocal}.
More specifically, even if signal and idler photons experience dispersion effects independently on their own path, the degree of coincidence can be maintained as if there are no dispersive media.
Thus, it has a great promise in resolving entanglement loss that significantly degrades the system performance of quantum communication technology.
Recently, several experimental works have been performed \cite{Baek2009nonlocal} to validate the NLDC effect even for few tens of kilometers \cite{PhysRevA.100.053803}.
Here, for the first time, we conduct numerical experiments to confirm the NLDC effect via the proposed quantization scheme.

\begin{figure*}
\centering\includegraphics[width=1\linewidth]
{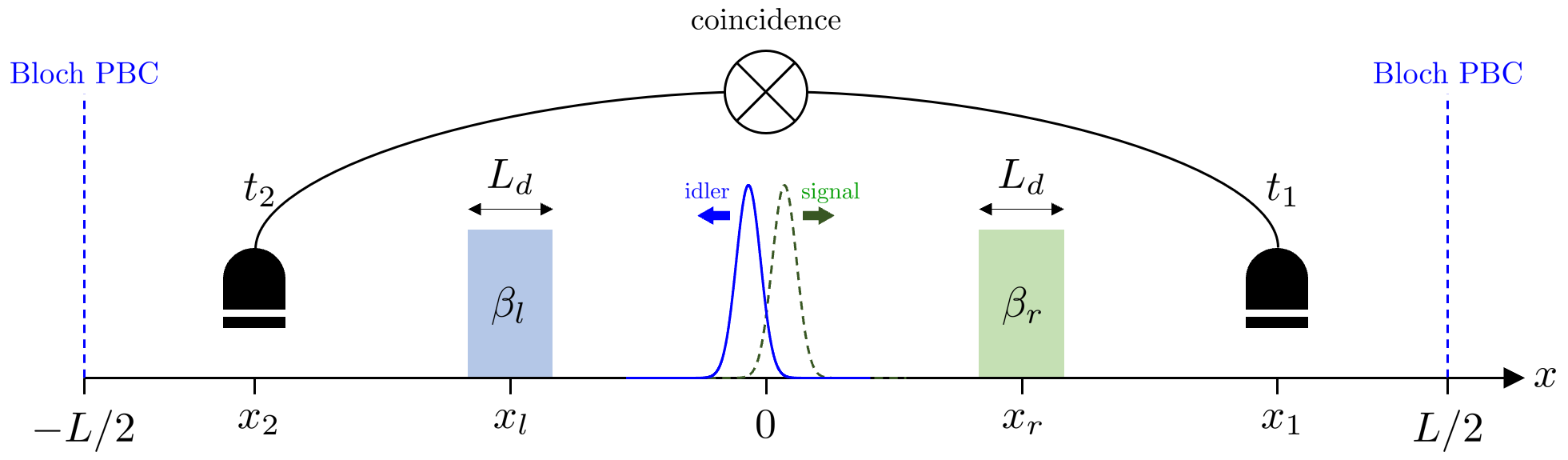}
\caption{
Problem geometry of 1-D simulations to observe non-local dispersion cancellation.
An energy-time entangled (or non-entangled) photon pair is initialized at $x=0$ and the signal and idler photons are supposed to propagate to the right and left sides, respectively.
We place a dispersive medium on each photon's path, denoted by $\beta_{r}$ and $\beta_{l}$ where $\beta$ represents a second-order dispersion of the medium.
The degree of coincidence is computed from two photodetections at $(x_{1},t_1)$ and $(x_{2},t_2)$.
}
\label{fig:NLDC_problem_geometry}
\end{figure*}

Again, we consider a 1-D problem geometry, as illustrated in Fig. \ref{fig:NLDC_problem_geometry}.
An energy-time entangled photon pair is initialized at $x=0$ and signal and idler photons are supposed to propagate to the right and left sides, respectively.
Again, each photon is polychromatic.
We place a dispersive medium on each photon's path, denoted by $\beta_{r}$ and $\beta_{l}$ where $\beta$ represents a second-order dispersion of the medium.
The degree of coincidence is computed from two photodetections at $(x_{1},t_1)$ and $(x_{2},t_2)$.
We present simulation parameters, numerical setup, design of dispersive media, and modeling the energy-time entangled photons in Appendix \ref{App_2g} in detail.
\begin{figure*}
\centering
\subfloat[entangled photon pair]{\includegraphics[width=.5\linewidth]{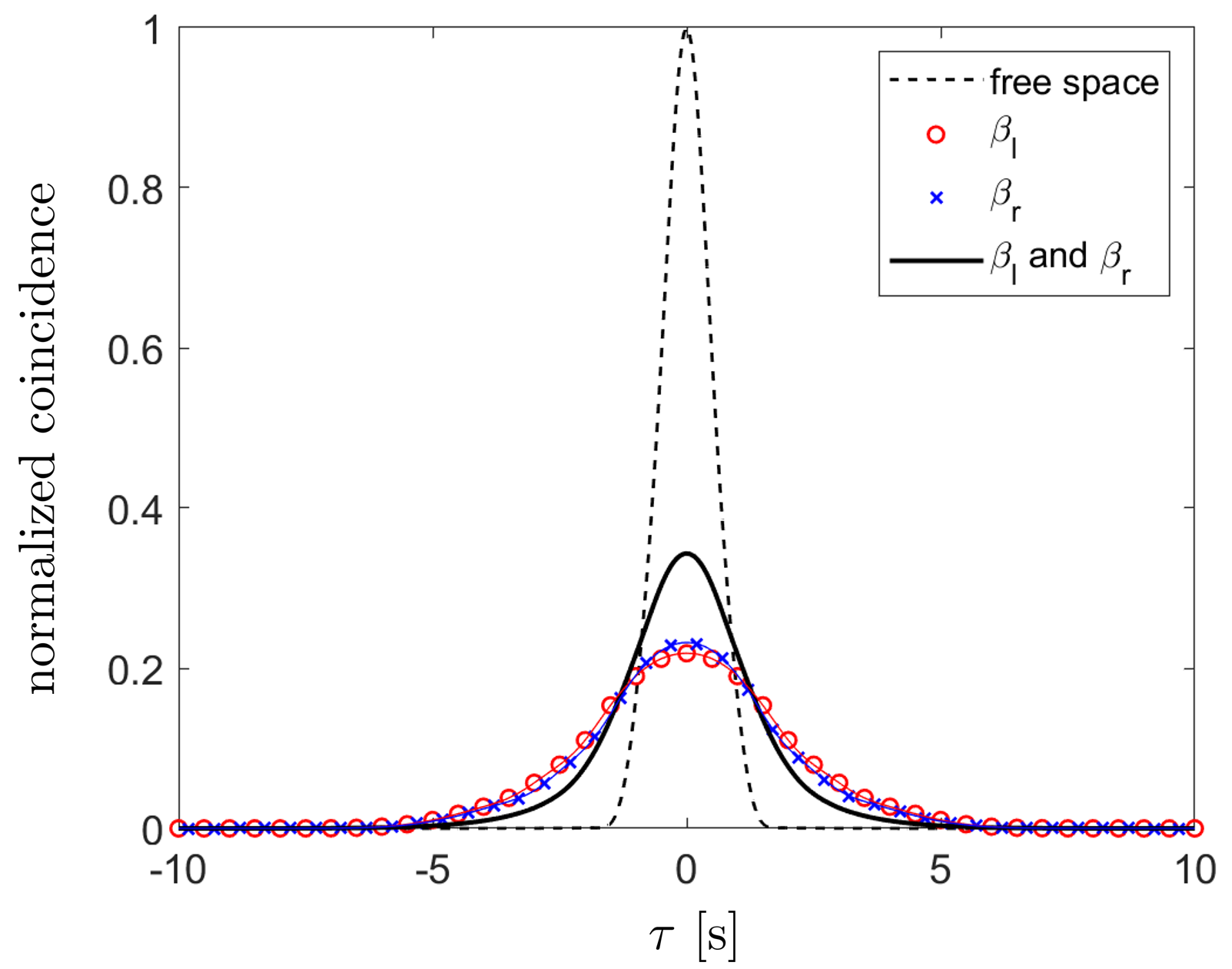}}
\subfloat[non-entangled photon pair]{\includegraphics[width=.5\linewidth]
{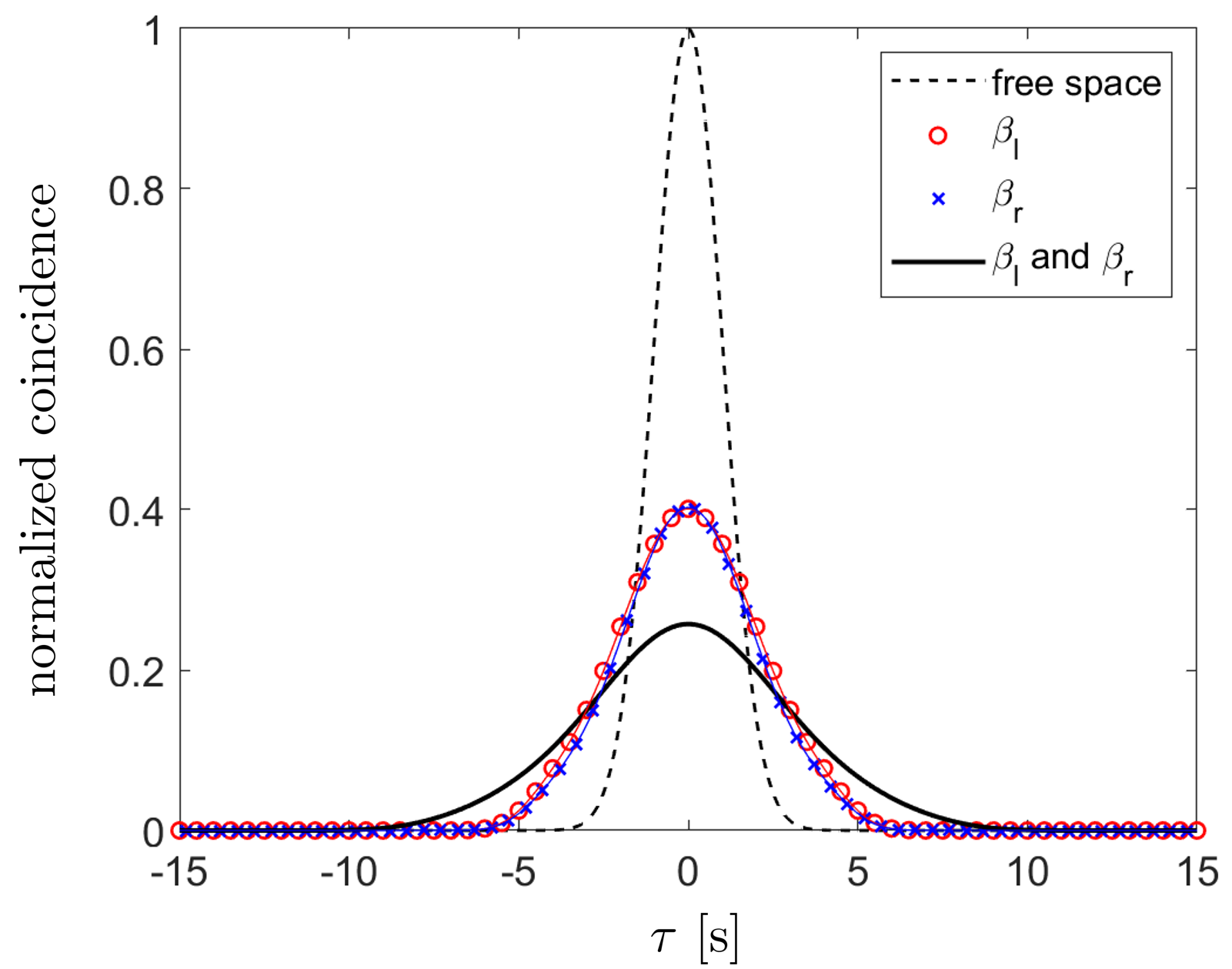}}
\caption{
Coincidence versus time difference $\tau$, defined by $\tau=(t_1-\tilde{t}_1)-(t_2-\tilde{t}_2)$ where $\tilde{t}_{i}$ is the delay of arrival time of $i$-th photon to photodetector.
}
\label{fig:NCI_ETET_NET}
\end{figure*}
Note that we again use the FDM with Bloch-Floquet boundary conditions to solve \eqref{eqn:GHEVP_no_cross_coupling}.

We conducted four simulations for both entangled and non-entangled photons: (1) free space (no dispersive media), (2) in the presence of left dispersive media ($\beta_{l}$), (3) right dispersive media ($\beta_{r}$), and (4) both dispersive media ($\beta_{l}$ and $\beta_{r}$). Then, we computed two-time coincidences for all cases.
Fig. \ref{fig:NCI_ETET_NET} displays aggregate coincidence (normalized by the free space case) versus time difference $\tau$, defined by $\tau=\left(t_{1}-\tilde{t}_{1}\right)-\left(t_{2}-\tilde{t}_{2}\right)$.
Here, $\tilde{t}_{i}$ denotes a delay of $i$-th photon arrival time to $i$-th photodetector for $i=1,2$, compared with the free space case.
Note that the delay is mostly affected by the degree of the first-order dispersion $\alpha$ \cite{Baek2009nonlocal}.

For the entangled photon pair in Fig. \ref{fig:NCI_ETET_NET}(a), the presence of either left or right dispersive media gets the coincidence peak broadened, compared with the free space case.
Furthermore, since both left and right dispersive media have the same magnitude of the second-order dispersion, the broadening amounts are almost equal.
It is very interesting to observe that when both dispersive media are present, the coincidence curve becomes narrower, resulting from the destructive interference of the second-order dispersion.
However, due to imperfect cancellations in higher-order dispersions, the coincidence peak is not perfectly converging to the free space case \cite{Ryu:17}.
On the other hand, as observed in Fig. \ref{fig:NCI_ETET_NET}(b), the non-entangled photon pair do not exhibit any dispersion cancellation that the coincidence curve in the presence of both dispersive media gets wider than one-sided medium cases.

\section{Conclusion}
We have presented a new mathematical modeling approach, called {\it canonical quantization with numerical mode-decomposition}, suited for studying how incoming photons interact with finite-sized dispersive media, which may not be simply described by the previous Fano-diagonalization-based quantization methods.
The main procedure was to (1) inspect a system where electromagnetic (EM) fields are coupled to non-uniformly-distributed Lorentz oscillators in Hamiltonian mechanics, (2) derive a generalized Hermitian eigenvalue problem for conjugate pairs on Euclidean space, (3) apply computational electromagnetics methods to find a countably-finite set of time-harmonic eigenmodes which diagonalizes the Hamiltonian, and (4) perform the subsequent canonical quantization with mode-decomposition.
Moreover, we have provided several numerical simulations for fully-quantum-theoretic phenomena, not predictable by classical Maxwell's equations, such as non-local dispersion cancellation of an entangled photon pair and Hong-Ou-Mandel (HOM) effect in a dispersive beam splitter, showing the great promise of the CEM-driven QEM/QO research.
We have shown the equivalence between the present approach and the recent works \cite{Jauslin2019canonical,Jauslin2020Critical} by the Jauslin's group, though, our formulation was based on Euclidean space with the use of CEM methods rather than Reciprocal space.

In the future, we will compare the computational efficiency of the present and Jauslin's group's formulations for various cases.
Furthermore, we will extend the present approach to dissipative quantum systems by introducing coarse-grained bath oscillators.
Moreover, we will investigate how to remove the redundancy when using Lorenz gauge, which is important to account for trapped modes, such as, surface plasmon polaritons.

\begin{acknowledgments}
We thank Dr. H. R. Jauslin for providing useful comments and discussions.
The work is funded by NSF 1818910 award and a startup fund at Purdue university.
WCC is also funded by DVSS at HKU, Summer 2019.
\end{acknowledgments}

\appendix
\section{Hamilton's EoMs in block matrix from}\label{App_1}
The Hamilton's EoMs can be expressed in the block matrix form as
\begin{flalign}
\frac{\partial}{\partial t}
\left[
\begin{matrix}
\mathbf{q} \\
\mathbf{p}
\end{matrix}
\right]
&=
\left[
\begin{matrix}
\frac{\delta H}{\delta \mathbf{p}} \\
-\frac{\delta H}{\delta \mathbf{q}}
\end{matrix}
\right]
=
\overline{\mathbf{J}}
\cdot
\left[
\begin{matrix}
\frac{1}{\delta  \mathbf{q}} \\
\frac{1}{\delta \mathbf{p}}
\end{matrix}
\right]
\delta H
\label{eqn_A_2}
\end{flalign}
where the differential Hamiltonian is given by
\begin{flalign}
\delta H
&=
\frac{1}{2}
\int_{V}d\mathbf{r}
\left[
\begin{matrix}
\delta \mathbf{q} \\
\delta \mathbf{p}
\end{matrix}
\right]^{\dag}
\cdot
\overline{\mathbf{U}}
\cdot
\left[
\begin{matrix}
\mathbf{q} \\
\mathbf{p}
\end{matrix}
\right]
+
\left[
\begin{matrix}
\mathbf{q} \\
\mathbf{p}
\end{matrix}
\right]^{\dag}
\cdot
\overline{\mathbf{U}}
\cdot
\left[
\begin{matrix}
\delta \mathbf{q} \\
\delta \mathbf{p}
\end{matrix}
\right]
\nonumber \\
&=
\int_{V}d\mathbf{r}
\left[
\begin{matrix}
\delta \mathbf{q} \\
\delta \mathbf{p}
\end{matrix}
\right]^{\dag}
\cdot
\overline{\mathbf{U}}
\cdot
\left[
\begin{matrix}
\mathbf{q} \\
\mathbf{p}
\end{matrix}
\right],
\nonumber \\
\overline{\mathbf{U}}
&=
\left[
\begin{matrix}
\overline{\mathbf{K}} & \overline{\mathbf{C}} \\
\overline{\mathbf{C}}^{\dag} & \overline{\mathbf{M}}
\end{matrix}
\right],
\quad
\overline{\mathbf{J}}
=
\left[
\begin{matrix}
\overline{\mathbf{0}} & \overline{\mathbf{I}}\\
-\overline{\mathbf{I}} & \overline{\mathbf{0}}
\end{matrix}
\right],
\end{flalign}
and $\overline{\mathbf{I}}$ is an identity matrix.

Substituting the above differential Hamiltonian into \eqref{eqn_A_2} yields
\begin{flalign}
\frac{\partial}{\partial t}
\left[
\begin{matrix}
\mathbf{q} \\
\mathbf{p}
\end{matrix}
\right]
&=
\overline{\mathbf{J}}
\cdot
\left[
\begin{matrix}
\frac{1}{\delta  \mathbf{q}} \\
\frac{1}{\delta \mathbf{p}}
\end{matrix}
\right]
\cdot
\left(
\int_{V}d\mathbf{r}
\left[
\begin{matrix}
\delta \mathbf{q} \\
\delta \mathbf{p}
\end{matrix}
\right]^{\dag}
\cdot
\overline{\mathbf{U}}
\cdot
\left[
\begin{matrix}
\mathbf{q} \\
\mathbf{p}
\end{matrix}
\right]
\right)
\nonumber \\
&=
\overline{\mathbf{J}}
\cdot
\left(
\int_{V}d\mathbf{r}
\left[
\begin{matrix}
\frac{1}{\delta  \mathbf{q}} \\
\frac{1}{\delta \mathbf{p}}
\end{matrix}
\right]
\cdot
\left[
\begin{matrix}
\delta \mathbf{q} \\
\delta \mathbf{p}
\end{matrix}
\right]^{\dag}
\cdot
\overline{\mathbf{U}}
\cdot
\left[
\begin{matrix}
\mathbf{q} \\
\mathbf{p}
\end{matrix}
\right]
\right).
\end{flalign}
One can make the use of the properties of functional derivatives \cite{haken1983quantum,CHEW2016quantum,lancaster2014quantum,stevens1995six}, as a result,
\begin{flalign}
\left[
\begin{matrix}
\frac{1}{\delta  \mathbf{q}(\mathbf{r}',t)} \\
\frac{1}{\delta \mathbf{p}(\mathbf{r}',t)}
\end{matrix}
\right]
\cdot
\left[
\begin{matrix}
\delta \mathbf{q}(\mathbf{r},t) \\
\delta \mathbf{p}(\mathbf{r},t)
\end{matrix}
\right]^{\dag}
&=
\left[
\begin{matrix}
\frac{\delta \mathbf{q}(\mathbf{r},t)}{\delta \mathbf{q}(\mathbf{r}',t)} &
\frac{\delta \mathbf{p}(\mathbf{r},t)}{\delta \mathbf{q}(\mathbf{r}',t)}
\\
\frac{\delta \mathbf{q}(\mathbf{r},t)}{\delta \mathbf{p}(\mathbf{r}',t)} &
\frac{\delta \mathbf{p}(\mathbf{r},t)}{\delta \mathbf{p}(\mathbf{r}',t)}
\end{matrix}
\right]
\nonumber \\
&=
\left[
\begin{matrix}
\overline{\boldsymbol{\delta}}(\mathbf{r}-\mathbf{r}') & 0 \\
0  & \overline{\boldsymbol{\delta}}(\mathbf{r}-\mathbf{r}')
\end{matrix}
\right].
\end{flalign}
Upon applying the sifting property of the delta function, the resulting Hamilton's EoMs become
\begin{flalign}
\frac{\partial}{\partial t}
\left[
\begin{matrix}
\mathbf{q} \\
\mathbf{p}
\end{matrix}
\right]
&=
\left(
\int_{V}d\mathbf{r}
\left[
\begin{matrix}
0 & \overline{\boldsymbol{\delta}}(\mathbf{r}-\mathbf{r}') \\
-\overline{\boldsymbol{\delta}}(\mathbf{r}-\mathbf{r}') & 0 \\
\end{matrix}
\right]
\cdot
\overline{\mathbf{U}}
\cdot
\left[
\begin{matrix}
\mathbf{q} \\
\mathbf{p}
\end{matrix}
\right]
\right)
\nonumber \\
&=
\overline{\mathbf{J}}
\cdot
\overline{\mathbf{U}}
\cdot
\left[
\begin{matrix}
\mathbf{q} \\
\mathbf{p}
\end{matrix}
\right]
=
\left[
\begin{matrix}
\overline{\mathbf{C}}^{\dag} & \overline{\mathbf{M}} \\
- \overline{\mathbf{K}} &  -\overline{\mathbf{C}}
\end{matrix}
\right]
\cdot
\left[
\begin{matrix}
\mathbf{q} \\
\mathbf{p}
\end{matrix}
\right].
\end{flalign}
The above procedure was also applied to arriving at \eqref{eqn:mc_EoMs}.

\section{Non-Hermicity of EVP due to cross-coupling terms}\label{App_2b}
Let us represent the dynamical variables by the linear superposition of time-harmonic eigenmodes as
\begin{flalign}
\left[
\begin{matrix}
\mathbf{q} \\
\mathbf{p}
\end{matrix}
\right]
=
\int_{\Omega}d\omega
\sum_{\lambda}
\Bigl(
\left[
\begin{matrix}
\tilde{\mathbf{q}}_{\omega,\lambda}(\mathbf{r}) \\
\tilde{\mathbf{p}}_{\omega,\lambda}(\mathbf{r})
\end{matrix}
\right]
\underbrace{
c_{\omega,\lambda}
e^{-i\omega t}
}_{c_{\omega,\lambda}(t)}
\Bigr)
\label{eqn:mode_decomp}
\end{flalign}
where $\Omega$ denotes a set including both positive and negative eigenfrequencies.
By substituting \eqref{eqn:mode_decomp} into \eqref{eqn:H_EoMs_1} and replacing the time derivative by $-i\omega$, one arrives at
\begin{flalign}
\omega
\left[
\begin{matrix}
\tilde{\mathbf{q}}_{\omega,\lambda} \\
\tilde{\mathbf{p}}_{\omega,\lambda}
\end{matrix}
\right]
&=
i\overline{\mathbf{J}} \cdot \overline{\mathbf{U}}
\cdot
\left[
\begin{matrix}
\tilde{\mathbf{q}}_{\omega,\lambda} \\
\tilde{\mathbf{p}}_{\omega,\lambda}
\end{matrix}
\right].
\label{eqn:EVP_1}
\end{flalign}
However, since $i\overline{\mathbf{J}} \cdot \overline{\mathbf{U}}$ is a non-Hermitian matrix, solutions of  \eqref{eqn:EVP_1} may not ensure the completeness of time-harmonic eigenmodes nor realness of eigenfrequencies.

\section{Equivalence between the present formulation and \cite{Jauslin2019canonical}}\label{App_2c}
From the no-cross-coupling description, we can apply the canonical transformation for generalized position and momentum as
\begin{flalign}
\mathbf{q}'=\overline{\mathbf{M}}^{-\frac{1}{2}}\cdot\mathbf{q},\quad
\mathbf{p}'=\overline{\mathbf{M}}^{\frac{1}{2}}\cdot\mathbf{p}.
\end{flalign}
Then, the Hamiltonian \eqref{eqn:d_Hamil} can be brought to the form
\begin{flalign}
H
&=
\frac{1}{2}
\int_{V}d\mathbf{r}
\left[
\begin{matrix}
\mathbf{q} \\
\mathbf{p}
\end{matrix}
\right]^{\dag}
\cdot
\left[
\begin{matrix}
\overline{\mathbf{K}} & \overline{\mathbf{0}} \\
\overline{\mathbf{0}} & \overline{\mathbf{M}}
\end{matrix}
\right]
\cdot
\left[
\begin{matrix}
\mathbf{q} \\
\mathbf{p}
\end{matrix}
\right]
\nonumber \\
&=
\frac{1}{2}
\int_{V}d\mathbf{r}
\left[
\begin{matrix}
\mathbf{q}' \\
\mathbf{p}'
\end{matrix}
\right]^{\dag}
\cdot
\left[
\begin{matrix}
\overline{\mathbf{M}}^{\frac{1}{2}}\cdot\overline{\mathbf{K}}\cdot\overline{\mathbf{M}}^{\frac{1}{2}} & \overline{\mathbf{0}} \\
\overline{\mathbf{0}} & \overline{\mathbf{I}}
\end{matrix}
\right]
\cdot
\left[
\begin{matrix}
\mathbf{q}' \\
\mathbf{p}'
\end{matrix}
\right]
\nonumber \\
&=
\frac{1}{2}
\int_{V}d\mathbf{r}
\left(\mathbf{p}'\right)^{\dag}\cdot\mathbf{p}'
+
\left(\mathbf{q}'\right)^{\dag}\cdot \overline{\boldsymbol{\Omega}}^{2}\cdot\mathbf{q}'
\end{flalign}
where the positive symmetric operator $\overline{\boldsymbol{\Omega}}^{2}=\overline{\mathbf{M}}^{\frac{1}{2}}\cdot\overline{\mathbf{K}}\cdot\overline{\mathbf{M}}^{\frac{1}{2}}$, called {\it frequency operator}.
Next, we show the equivalence between the GH-EVP \eqref{eqn:GHEVP_no_cross_coupling} and equation (14) in \cite{Jauslin2019canonical}.
The standard eigenvalue problem of equation (14) in \cite{Jauslin2019canonical} is given by
\begin{flalign}
\overline{\mathbf{\Omega}}^{2}
\cdot \boldsymbol{\psi}_{\omega,\lambda}
&=
\omega^{2}
\boldsymbol{\psi}_{\omega,\lambda}
\nonumber \\
\overline{\mathbf{M}}^{\frac{1}{2}}\cdot\overline{\mathbf{K}}\cdot\overline{\mathbf{M}}^{\frac{1}{2}}
\cdot \boldsymbol{\psi}_{\omega,\lambda}
&=
\omega^{2}
\boldsymbol{\psi}_{\omega,\lambda}
\nonumber \\
\overline{\mathbf{K}}\cdot\overline{\mathbf{M}}^{\frac{1}{2}}
\cdot \boldsymbol{\psi}_{\omega,\lambda}
&=
\omega^{2}
\overline{\mathbf{M}}^{-\frac{1}{2}}\cdot\boldsymbol{\psi}_{\omega,\lambda}.
\end{flalign}
Identifying $\tilde{\mathbf{q}}_{\omega,\lambda}=\overline{\mathbf{M}}^{\frac{1}{2}}
\cdot \boldsymbol{\psi}_{\omega,\lambda}$, the above can be written by
\begin{flalign}
\overline{\mathbf{K}}\cdot\tilde{\mathbf{q}}_{\omega,\lambda}
&=
\omega^{2}
\overline{\mathbf{M}}^{-1}\cdot\tilde{\mathbf{q}}_{\omega,\lambda}.
\end{flalign}
Thus, our GH-EVP and that in \cite{Jauslin2019canonical} are mathematically equivalent.
It is to be noted that our GH-EVP is based on Euclidean space, though, equation (14) in \cite{Jauslin2019canonical} is based on Reciprocal space.

\section{Diagonalization of the classical Hamiltonian}\label{App_2d}
The generalized momentum variable can be represented by
\begin{flalign}
\mathbf{p}=\int_{\Omega_{+}}d
\omega \sum_{\lambda}
\tilde{\mathbf{p}}_{\omega,\lambda}d_{\omega,\lambda}e^{-i\omega t}+\text{h.c.}
\label{eqn:p_no_coupling_1}
\end{flalign}
By substituting \eqref{eqn:q_no_coupling} and \eqref{eqn:p_no_coupling_1} into the original Hamiltonian \eqref{eqn:d_Hamil} and rearranging it, one can arrive at
\begin{flalign}
H
&=
\frac{1}{2}
\int_{V}d\mathbf{r}
\Biggl(
\int_{\Omega_{+}}
d\omega
\sum_{\lambda}
\tilde{\mathbf{q}}_{\omega,\lambda}^{\dag}
d^{*}_{\omega,\lambda}
e^{i\omega t}
\Biggr)
\nonumber \\
&
~~~~~~~\cdot\overline{\mathbf{K}}
\cdot
\Biggl(
\int_{\Omega_{+}}
d\omega'
\sum_{\lambda'}
\tilde{\mathbf{q}}_{\omega',\lambda'}
d_{\omega',\lambda'}
e^{-i\omega' t}
\Biggr)
\nonumber \\
&+
\frac{1}{2}
\int_{V}d\mathbf{r}
\Biggl(
\int_{\Omega_{+}}
d\omega
\sum_{\lambda}
\tilde{\mathbf{p}}_{\omega,\lambda}^{\dag}
d^{*}_{\omega,\lambda}
e^{i\omega t}
\Biggr)
\nonumber \\
&
~~~~~~~\cdot\overline{\mathbf{M}}
\cdot
\Biggl(
\int_{\Omega_{+}}
d\omega'
\sum_{\lambda'}
\tilde{\mathbf{p}}_{\omega',\lambda'}
d_{\omega',\lambda'}
e^{-i\omega' t}
\Biggr)
\nonumber \\
&=
\frac{1}{2}
\int_{\Omega_{+}}d\omega
\sum_{\lambda}
\int_{\Omega_{+}}d\omega'
\sum_{\lambda'}
d^{*}_{\omega,\lambda}
d_{\omega',\lambda'}
e^{i(\omega-\omega') t}
\nonumber \\
&
~~~~~~~\times
\Biggl(
\int_{V}d\mathbf{r}~
\tilde{\mathbf{q}}_{\omega,\lambda}^{\dag}
\cdot
\overline{\mathbf{K}}
\cdot
\tilde{\mathbf{q}}_{\omega',\lambda'}
\Biggr)
\nonumber \\
&+
\frac{1}{2}
\int_{\Omega_{+}}d\omega
\sum_{\lambda}
\int_{\Omega_{+}}d\omega'
\sum_{\lambda'}
d^{*}_{\omega,\lambda}
d_{\omega',\lambda'}
e^{i(\omega-\omega') t}
\nonumber \\
&
~~~~~~~\times
\Biggl(
\int_{V}d\mathbf{r}~
\tilde{\mathbf{p}}_{\omega,\lambda}^{\dag}
\cdot
\overline{\mathbf{M}}
\cdot
\tilde{\mathbf{p}}_{\omega',\lambda'}
\Biggr).
\label{eqn:H_re_d}
\end{flalign}
Using the relation in the first row equation of \eqref{eqn:mc_EoMs}, a time-harmonic eigenmode for $\mathbf{p}$ can be represented by
\begin{flalign}
\tilde{\mathbf{p}}_{\omega,\lambda}
=
-i\omega\overline{\mathbf{M}}^{-1}\cdot\tilde{\mathbf{q}}_{\omega,\lambda},
\end{flalign}
therefore, one can have the following property
\begin{flalign}
&\int_{V}d\mathbf{r} ~
\tilde{\mathbf{p}}_{\omega,\lambda}^{\dag}
\cdot
\overline{\mathbf{M}}
\cdot
\tilde{\mathbf{p}}_{\omega',\lambda'}
\nonumber \\
&=
\omega\omega'
\int_{V}d\mathbf{r}
\left(
\overline{\mathbf{M}}^{-1}\cdot\tilde{\mathbf{q}}_{\omega,\lambda}
\right)^{\dag}
\cdot
\overline{\mathbf{M}}
\cdot
\left(
\overline{\mathbf{M}}^{-1}\cdot\tilde{\mathbf{q}}_{\omega',\lambda'}
\right)
\nonumber \\
&=
\omega\omega'
\int_{V}d\mathbf{r} ~
\tilde{\mathbf{q}}_{\omega,\lambda}^{\dag}
\cdot
\overline{\mathbf{M}}^{-1}
\cdot
\tilde{\mathbf{q}}_{\omega',\lambda'}
=
\omega\omega'\delta_{\omega,\omega'}\delta_{\lambda,\lambda'}.
\label{eqn:p_property}
\end{flalign}
Finally, by applying \eqref{eqn:OT_2_second_app} and \eqref{eqn:p_property} to \eqref{eqn:H_re_d}, one can obtain the diagonalized Hamiltonian such as
\begin{flalign}
H
&=
\frac{1}{2}
\int_{\Omega_{+}}d\omega\sum_{\lambda}
\omega^{2}
\Bigl(
d^{*}_{\omega,\lambda}
d_{\omega,\lambda}
+
d_{\omega,\lambda}
d^{*}_{\omega,\lambda}
\Bigr).
\end{flalign}

\section{Cross-coupling description}\label{App_2e}
Motivated by the recent work to rigorously find time-harmonic eigenmodes for photonic crystal systems by solving an explicit EVP \cite{Raman2010Photonic}, we can derive another GH-EVP, called {\it cross-coupling description}.
To convert \eqref{eqn:EVP_1} into a GH-EVP, as performed in \cite{Raman2010Photonic}, we multiply $\overline{\mathbf{U}}$ to the both sides of \eqref{eqn:EVP_1} to obtain
\begin{flalign}
\boxed{
\omega
\overline{\mathbf{U}}
\cdot
\left[
\begin{matrix}
\tilde{\mathbf{q}}_{\omega,\lambda} \\
\tilde{\mathbf{p}}_{\omega,\lambda}
\end{matrix}
\right]
=
i
\overline{\mathbf{V}}
\cdot
\left[
\begin{matrix}
\tilde{\mathbf{q}}_{\omega,\lambda} \\
\tilde{\mathbf{p}}_{\omega,\lambda}
\end{matrix}
\right]
}
\label{eqn:GEVP_final}
\end{flalign}
where $i\overline{\mathbf{V}}=\overline{\mathbf{U}} \cdot i \overline{\mathbf{J}} \cdot \overline{\mathbf{U}}$ is now a Hermitian matrix.
Finally, solving \eqref{eqn:GEVP_final} yields a complete set of time-harmonic eigenmodes with real eigenfrequencies $\omega$.
Furthermore, following two orthonormal properties can be deduced
\begin{flalign}
\int_{V}d\mathbf{r}
\Bigl(
\left[
\begin{matrix}
\tilde{\mathbf{q}}_{\omega,\lambda} \\
\tilde{\mathbf{p}}_{\omega,\lambda}
\end{matrix}
\right]^{\dag}
\cdot
\overline{\mathbf{U}}
\cdot
\left[
\begin{matrix}
\tilde{\mathbf{q}}_{\omega',\lambda'} \\
\tilde{\mathbf{p}}_{\omega',\lambda'}
\end{matrix}
\right]
\Bigr)
&=
\delta_{\omega,\omega'}\delta_{\lambda,\lambda'},
\label{eqn:OT1}
\\
\int_{V}d\mathbf{r}
\Bigl(
\left[
\begin{matrix}
\tilde{\mathbf{q}}_{\omega,\lambda} \\
\tilde{\mathbf{p}}_{\omega,\lambda}
\end{matrix}
\right]^{\dag}
\cdot
i
\overline{\mathbf{V}}
\cdot
\left[
\begin{matrix}
\tilde{\mathbf{q}}_{\omega',\lambda'} \\
\tilde{\mathbf{p}}_{\omega',\lambda'}
\end{matrix}
\right]
\Bigr)
&=
\omega\delta_{\omega,\omega'}\delta_{\lambda,\lambda'}.
\label{eqn:OT2}
\end{flalign}
Substituting \eqref{eqn:mode_decomp} into the original Hamiltonian \eqref{eqn:classical_Ham} and applying the orthonormal condition \eqref{eqn:OT2}, one can easily diagonalize the Hamiltonian in terms of $c_{\omega,\lambda}$ as
\begin{flalign}
\boxed{
H
=
\frac{1}{2}
\int_{\Omega_{+}}
d\omega
\sum_{\lambda}
\Bigl(
c^{*}_{\omega,\lambda}
c_{\omega,\lambda}
+
c_{\omega,\lambda}
c^{*}_{\omega,\lambda}
\Bigr)
}
\label{eqn:H_diag_PC_coupling}
\end{flalign}
where $\Omega_{+}$ denotes the positive frequency regime of $\Omega$.

The manipulation done to derive \eqref{eqn:GEVP_final} is more than a coincidence.
In fact, \eqref{eqn:GEVP_final} is closely associated with the energy continuity equation.
To check this, let us multiply $\overline{\mathbf{U}}$ to the both sides of \eqref{eqn:H_EoMs_1} as
\begin{flalign}
\overline{\mathbf{U}}
\cdot
\Biggl(
\frac{\partial}{\partial t}
\left[
\begin{matrix}
\mathbf{q} \\
\mathbf{p}
\end{matrix}
\right]
\Biggr)
&=
\overline{\mathbf{V}}
\cdot
\left[
\begin{matrix}
\mathbf{q} \\
\mathbf{p}
\end{matrix}
\right]
\label{eqn:TD_GEVP}
\end{flalign}
which is the time-domain description of \eqref{eqn:GEVP_final}.
It should be mentioned that $\overline{\mathbf{U}}$ and $\overline{\mathbf{V}}$ are real matrices and independent of time or frequency.
Then multiplying $\left[\mathbf{q},~\mathbf{p}\right]^{*}$ to the both sides of \eqref{eqn:TD_GEVP}, one arrives at
\begin{flalign}
\left[
\begin{matrix}
\mathbf{q} \\
\mathbf{p}
\end{matrix}
\right]^{\dag}
\cdot
\overline{\mathbf{U}}
\cdot
\Biggl(
\frac{\partial}{\partial t}
\left[
\begin{matrix}
\mathbf{q} \\
\mathbf{p}
\end{matrix}
\right]
\Biggr)
&=
\left[
\begin{matrix}
\mathbf{q} \\
\mathbf{p}
\end{matrix}
\right]^{\dag}
\cdot
\overline{\mathbf{V}}
\cdot
\left[
\begin{matrix}
\mathbf{q} \\
\mathbf{p}
\end{matrix}
\right].
\label{eqn:energy_CE_gen}
\end{flalign}
The LHS in \eqref{eqn:energy_CE_gen} represents an energy density rate over an infinitesimal volume since
\begin{flalign}
\left[
\begin{matrix}
\mathbf{q} \\
\mathbf{p}
\end{matrix}
\right]^{\dag}
\cdot
\overline{\mathbf{U}}
\cdot
\Biggl(
\frac{\partial}{\partial t}
\left[
\begin{matrix}
\mathbf{q} \\
\mathbf{p}
\end{matrix}
\right]
\Biggr)
&=
2\frac{\partial}{\partial t}
\mathcal{H}
-
\Biggl(
\frac{\partial}{\partial t}
\left[
\begin{matrix}
\mathbf{q} \\
\mathbf{p}
\end{matrix}
\right]
\Biggr)^{\dag}
\cdot
\overline{\mathbf{U}}
\cdot
\left[
\begin{matrix}
\mathbf{q} \\
\mathbf{p}
\end{matrix}
\right]
\nonumber \\
&=
\frac{\partial}{\partial t}
\mathcal{H}
\end{flalign}
where
\begin{flalign}
\mathcal{H}=
\frac{1}{2}\left[
\begin{matrix}
\mathbf{q} \\
\mathbf{p}
\end{matrix}
\right]^{\dag}
\cdot
\overline{\mathbf{U}}
\cdot
\left[
\begin{matrix}
\mathbf{q} \\
\mathbf{p}
\end{matrix}
\right].
\end{flalign}
The energy continuity equation states that an energy density rate should be equal to a negative of an energy flux, viz.,
\begin{flalign}
\frac{\partial}{\partial t}\mathcal{H}
+
\text{energy flux}=0.
\end{flalign}
Thus, the RHS in \eqref{eqn:energy_CE_gen} can be interpreted as a negative of an energy flux flowing out of a closed surface of the infinitesimal volume
\begin{flalign}
\text{energy flux}
&=
-
\left[
\begin{matrix}
\mathbf{q} \\
\mathbf{p}
\end{matrix}
\right]^{\dag}
\cdot
\overline{\mathbf{V}}
\cdot
\left[
\begin{matrix}
\mathbf{q} \\
\mathbf{p}
\end{matrix}
\right]
.
\label{eqn:energy_flux}
\end{flalign}
One can easily check that the above energy flux only contains EM-associated terms ($\mathbf{A}$, $\Phi$, $\boldsymbol{\Pi}_{AP}$, and $\Pi_{\Phi}$) while $\mathbf{P}$ and $\boldsymbol{\Pi}_{P}$ terms are canceled out.
This coincides with the explanation in \cite{Huttner_1991} that Lorentz oscillators do not propagate energy.
Furthermore, when polarization density goes to zero, the energy continuity equation \eqref{eqn:energy_CE_gen} converges to the conventional Poynting theorem.

\begin{figure*}
\centering
\subfloat[overall coupling relationship]
{\includegraphics[width=.2825\linewidth]{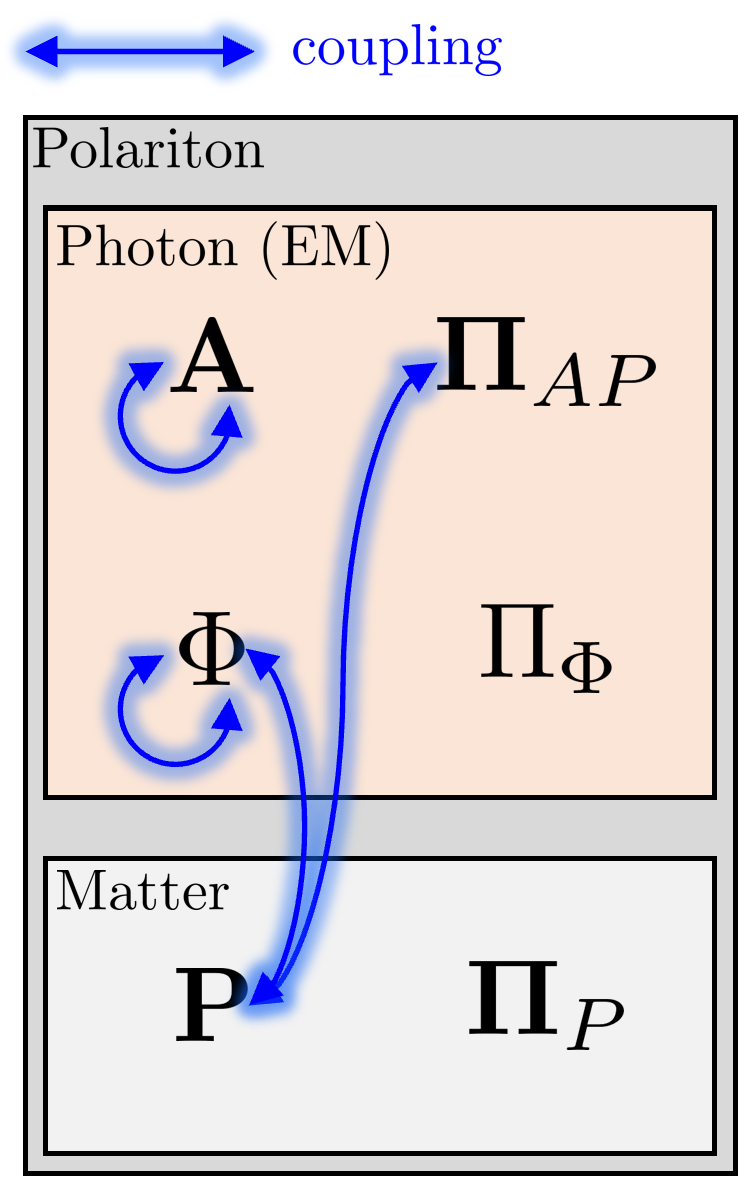}}
\quad\subfloat[two possible methods to derive generalized Hermitian eigenvalue problems]
{\includegraphics[width=.62\linewidth]{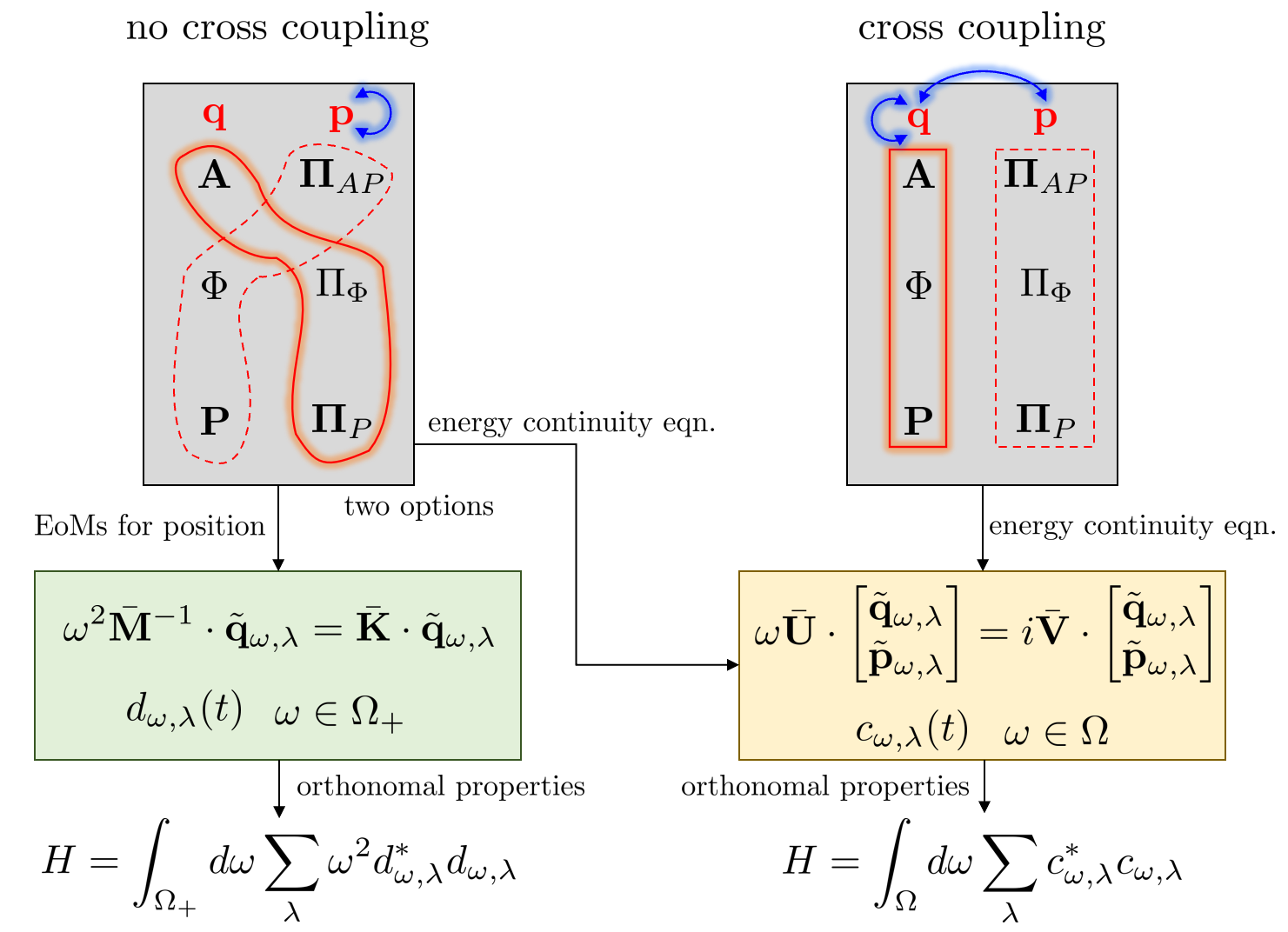}}
\caption{
Overview of the two possible ways to derive GH-EVPs for a coupled system between EM fields and lossless Lorentz oscillators.
The schematic in (a) describes an overall coupling relation among all dynamical variables in the original Hamiltonian \eqref{eqn:original_H}.
Note that blue solid-glowed line symbolizes coupling between two variables.
In (b), the left and right schematics depict the no cross coupling and cross coupling descriptions, respectively.
In the presence of cross coupling, one should invoke the energy continuity equation to obtain a GH-EVP which yields a full set of time-harmonic eigenmodes for both $\mathbf{q}$ and $\mathbf{p}$ with positive and negative eigenfrequencies.
In contrast, by properly defining $\mathbf{q}$ and $\mathbf{p}$ for no cross coupling, one can either arrive at another GH-EVP in terms of $\mathbf{q}$ only.
Consequently, it yields the smaller eigenspace spanned by time-harmonic eigenmodes for $\mathbf{q}$ with positive eigenfrequency.
Both methods can easily diagonalize the Hamiltonian in term of modal amplitudes either $c_{\omega,\lambda}$ or $d_{\omega,\lambda}$ via the orthonormal properties inherent from GH-EVPs.
}
\label{fig:description}
\end{figure*}

We have shown two possible approaches to derive GH-EVPs for EM fields coupled to lossless Lorentz oscillators and search for a complete set of time-harmonic eigenmodes of the system.
The overview is illustrated in Fig. \ref{fig:description}.

\section{Details of simulations on the Hong-Ou-Mandel effect in a dispersive beam splitter}\label{App_2f}

\begin{table}
\caption{Simulation parameters.}
\centering
\begin{tabular}{clclcl}
\toprule
\cmidrule(lr){1-2}\cmidrule(lr){3-4}\cmidrule(lr){5-6}
$L$ & $1.5$ [m] &
$N^{(0)}$ & $2,500$ &
$x_{g}$ & $0.3747$ [m]
\\
$L_{s}$ & $6$ [mm] &
$\Delta x$ & $0.6$ [mm] &
$\sigma_{g}$ & $0.05$ [m]
\\
$\epsilon_{s,\infty}$ & $7$ &
$N^{(0)}_s$ & $10$ &
$\omega_{g}$ & $526c$ [rad/s]
\\
\bottomrule
\end{tabular}
\label{tab:parameter}
\end{table}

\subsection{Design of dispersive beam splitter}
\begin{figure}
\centering
\includegraphics[width=\linewidth]
{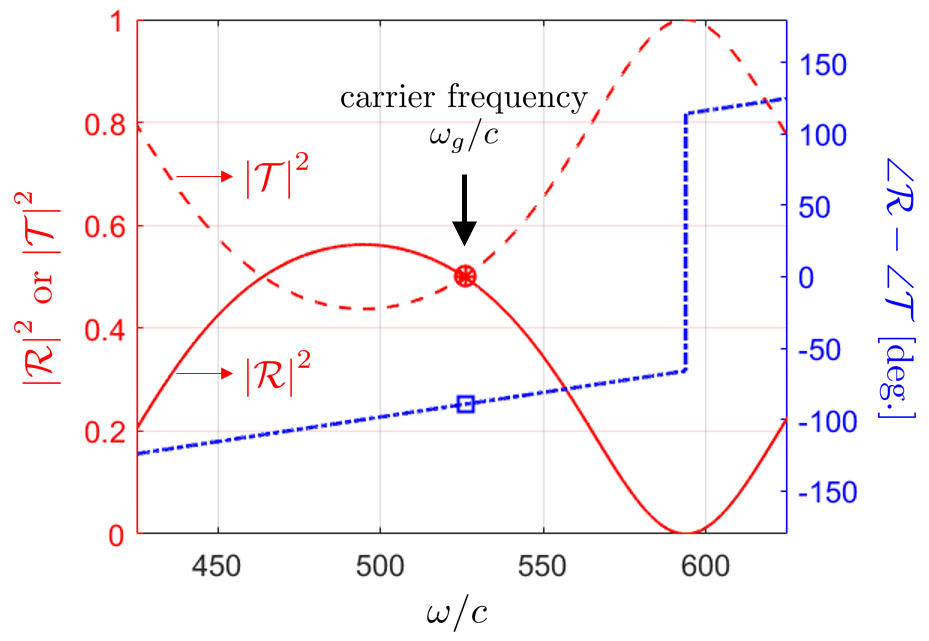}
\caption{
Reflectivity, transmissivity, and their phase difference versus $\omega$ for the designed dispersionless beam splitter.
The 50:50 reflectivity and transmissivity with a quadrature phase shift can be achieved around $\omega/c\approx 526$ which will be chosen for the center frequency of incident photons' wavepackets.
}
\label{fig:design_BS}
\end{figure}

A dispersionless beam splitter is designed first.
Performing a parametric study, we set the relative permittivity of the slab $\epsilon_{s,\infty}=7\epsilon_{0}$ [F/m] and the thickness $L_{s}=6$ [mm].
Relevant parameters are listed in Table \ref{tab:parameter}.
It can be observed in Fig. \ref{fig:design_BS} that the 50:50 reflectivity $\left|\mathcal{R}\right|^{2}$ and transmissivity $\left|\mathcal{T}\right|^{2}$ with a quadrature phase shift can be achieved around $\omega/c\approx 526$.
This frequency $\omega/c\approx 526$ will be used for the carrier frequency $\omega_g$ of incident photons' wavepackets.
It is to be noted that even though material dispersion is ignored, geometrical dispersion is present due to the finite thickness of the beam splitter.

\begin{figure}
\centering
\includegraphics[width=\linewidth]
{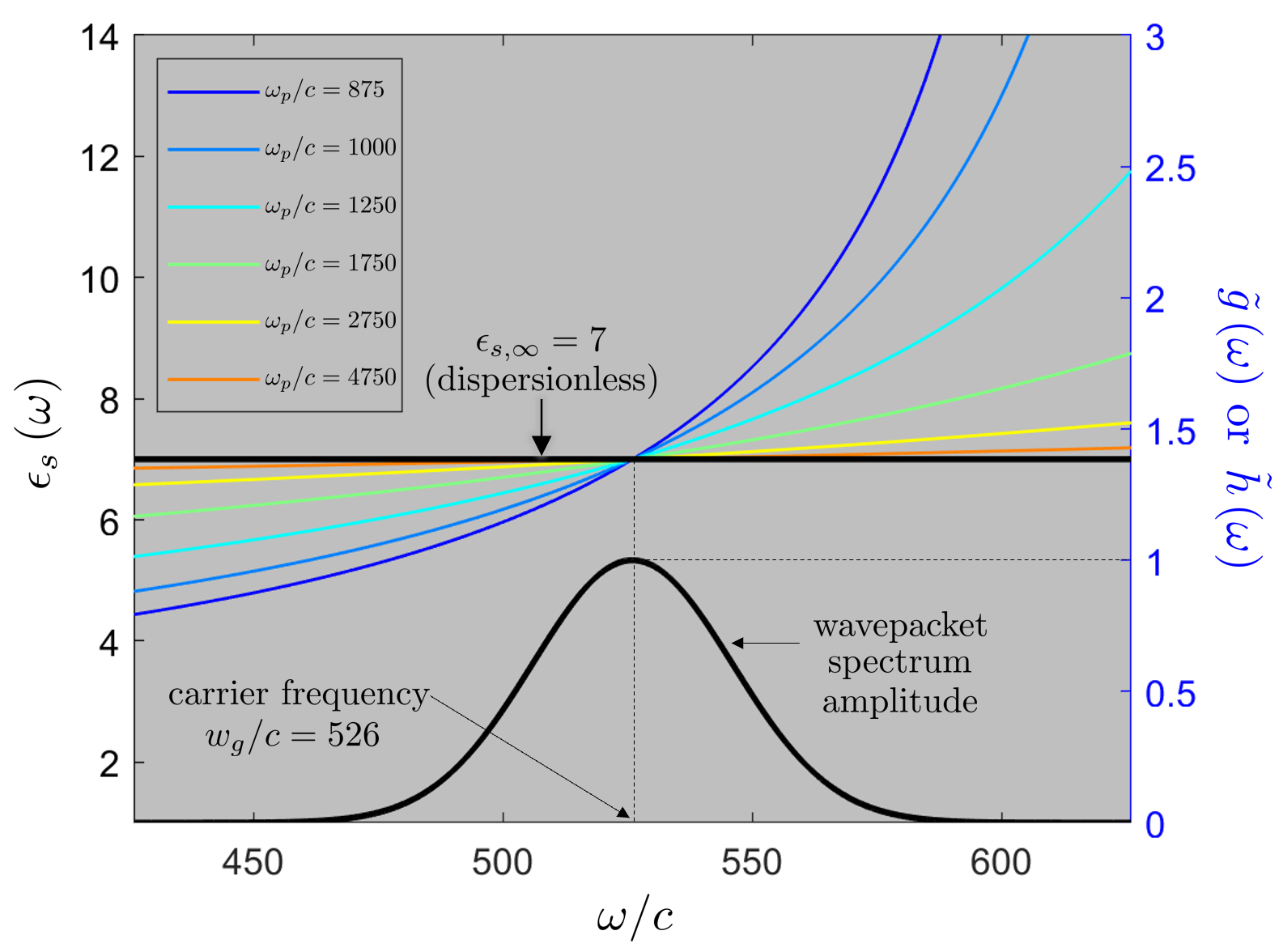}
\caption{
Relative dielectric constant $\epsilon_{s}(\omega)$ versus $\omega$ for various plasma frequencies.
In dispersive cases, since $\epsilon_{s}(\omega)$ starts deviating from $\epsilon_{s,\infty}=7\epsilon_0$ as $\left|\omega-526c\right|$ increases, the 50:50 performance bandwidth decreases.
The the smaller $\omega_p$ (higher dispersion) is, the smaller the bandwidth becomes.
}
\label{fig:dispersive_BS}
\end{figure}

The dispersive dielectric slab is modeled by single species Lorentz oscillators.
All Lorentz oscillators have same $\omega_{p}$ and $\omega_{0}$ where $\omega_{0}^{2}=\omega_{p}^{2}/(\epsilon_{s,\infty}-1)+\omega_{g}^{2}$.
Since the resulting relative dielectric constant becomes $\epsilon_{s}(\omega)=1+{\omega_{p}^{2}}/\left({\omega_{0}^{2}-\omega^{2}}\right)$, it always ensures $\epsilon_{s}(\omega_g)=\epsilon_{s,\infty}$.
Fig. \ref{fig:dispersive_BS} illustrates the dielectric constant versus $\omega$ for various $\omega_{p}$.
Because $\epsilon_{s}(\omega)$ starts deviating from $\epsilon_{s,\infty}=7\epsilon_0$ when $\left|\omega-w_g\right|$ increases, the bandwidth of exhibiting the 50:50 performance becomes narrower.
The the smaller $\omega_p$ (higher dispersion) is, the narrower the 50:50 performance bandwidth is.

\subsection{Extraction of numerical eigenmodes}
To extract numerical time-harmonic eigenmodes, we use the finite-difference mthod (FDM) and Bloch-Floquet boundary conditions to numerically solve \eqref{eqn:disc_GH_evp}.
The problem domain $V\in\left\{x\in\left[-L/2,L/2\right]\right\}$ is uniformly discretized (grid spacing $\Delta x$) by $N^{(0)}$ number of grid points.
The number of grid points inside the beam splitter is $N_s^{(0)}$.
Thus, we have total $N^{(0)}+N_s^{(0)}$ number of numerical time-harmonic eigenmodes having positive eigenfrequencies.

\subsection{Modeling incoming polychromatic photons}
Two photons are assumed to be polychromatic, viz., they are riding on wavepackets whose spatial distributions are modeled by $g(x)$ and $h(x)$, respectively.
The corresponding initial quantum state can be modeled by
\begin{flalign}
\ket{\Psi^{(2)}}
&=
\Bigl(
\int_{\Omega_{+}}d\omega
\sum_{\lambda}
\tilde{g}(\omega,\lambda)\hat{d}^{\dag}_{\omega,\lambda}
\Bigr)
\nonumber \\
&\times
\Bigl(
\int_{\Omega_{+}}d\omega'
\sum_{\lambda'}
\tilde{h}(\omega',\lambda')\hat{d}^{\dag}_{\omega',\lambda'}
\Bigr)
\ket{0}
\nonumber \\
&\approx
\Bigl(
\sum_{m}\tilde{g}_{m}\hat{d}^{\dag}_{m}
\Bigr)
\Bigl(
\sum_{n}\tilde{h}_{n}\hat{d}^{\dag}_{n}
\Bigr)
\ket{0}
\nonumber \\
&=
(\tilde{\mathbf{g}}^{T}\cdot\hat{\mathbf{c}})
(\tilde{\mathbf{h}}^{T}\cdot\hat{\mathbf{c}})
\ket{0}
\label{eqn:qs_two_photon}
\end{flalign}
where the second equality is the discrete counterpart of the first one, $m$ and $n$ are numerical time-harmonic eigenmode indices, and $[\tilde{\mathbf{g}}]_{m}=\tilde{g}_{m}$ and $[\tilde{\mathbf{h}}]_{n}=\tilde{h}_{n}$ are spectral probability amplitudes.
If wavepacket is modeled by Gaussian function,
\begin{flalign}
g(x)
&=
g_{0}e^{-\left(\frac{x-(x_{g}-\delta x_{g})}{\sqrt{2}\sigma_{g}}\right)^{2}}e^{-i k_{g} x},
\\
h(x)
&=
h_{0}e^{-\left(\frac{x+x_{g}}{\sqrt{2}\sigma_{g}}\right)^{2}}e^{i k_{g} x},
\end{flalign}
where carrier wavenumber $k_{g}=\omega_{g}/c$ and $g_{0}$ and $h_{0}$ are normalization constants.
By using the orthonormal properties of numerical time-harmonic eigenmodes, one can obtain $\tilde{\mathbf{g}}$ and $\tilde{\mathbf{h}}$.

\subsection{Calculation of second order correlation}
We can evaluate the second order correlation \cite{Glauber1963quantum,Mandel1995optical}
\begin{flalign}
g^{(2)}(\tau)
=
\frac
{
\mel
{\Psi^{(2)}}
{
\hat{\alpha}^{(+)}
\hat{\beta}^{(-)}
\hat{\beta}^{(+)}
\hat{\alpha}^{(+)}
}
{\Psi^{(2)}}
}
{
\mel
{\Psi^{(2)}}
{
\hat{\alpha}^{(-)}
\hat{\alpha}^{(+)}
}
{\Psi^{(2)}}
\mel
{\Psi^{(2)}}
{
\hat{\beta}^{(-)}
\hat{\beta}^{(+)}
}
{\Psi^{(2)}}
}
\end{flalign}
where $\tau=\delta x_{g}/c$ and
\begin{flalign}
\hat{\alpha}^{(\pm)}&=\hat{A}^{(\pm)}(x_{1},t_{1}),
\\
\hat{\beta}^{(\pm)}&=\hat{A}^{\pm}(x_{2},t_{1}+\tau),
\end{flalign}
$t_{1}=2x_{g}/c$.
The detailed calculation for $g^{(2)}(\tau)$ is explained in \cite{Na2020quantum}.

\section{Details of simulations on non-local dispersion cancellation for an energy-time entangled photon pair}\label{App_2g}
\subsection{Modeling energy-time entangled photon pair}
Based on a pump frequency $\Omega_P/c=35$, the signal and idler photons have the center frequencies $\Omega_{S}/c=37.5$ and $\Omega_{I}/c=32.5$ with bandwidth of $5c$.
The corresponding initial quantum state can be written by
\begin{flalign}
\ket{\Psi}
&=
\int_{-\infty}^{\infty}d\omega_{2}\sum_{\lambda_{2}}
\int_{-\infty}^{\infty}d\omega_{1}\sum_{\lambda_{1}}
\psi(\omega_{2},\lambda_{2},\omega_{1},\lambda_{1})
\nonumber \\
&
\times
\hat{d}^{\dag}_{\omega_{2},\lambda_{2}}\hat{d}^{\dag}_{\omega_{1},\lambda_{1}}\ket{0}.
\end{flalign}
where $\psi$ is a non-factorizable joint spectral probability amplitude.
On the other hand, one can describe a (spatially-localized) non-entangled photon pair by
\begin{flalign}
\ket{\Phi}
&=
\int_{-\infty}^{\infty}d\omega_{2}\sum_{\lambda_{2}}
\int_{-\infty}^{\infty}d\omega_{1}\sum_{\lambda_{1}}
\phi_{2}(\omega_{2},\lambda_{2})
\phi_{1}(\omega_{1},\lambda_{1})
\nonumber \\
&\times
\hat{d}_{\omega_{2},\lambda_{2}}
\hat{d}_{\omega_{1},\lambda_{1}}
\ket{0}
\end{flalign}
where $\phi_{i}$ describes a spectral probability amplitude of $i$-th photon for $i=1,2$.
By implicitly accounting for the degeneracy index, Fig. \ref{fig:spectrum_coincidence_ETET_NET}(a) and Fig. \ref{fig:spectrum_coincidence_ETET_NET}(b) illustrate $\psi(\omega_{2},\omega_{1})$ and $\phi(\omega_{2},\omega_{1})=\phi_{2}(\omega_{2})\phi_{1}(\omega_{1})$.

\subsection{Design of dispersive media}
\begin{figure}
\centering
\subfloat[dispersion diagram]{\includegraphics[width=1\linewidth]{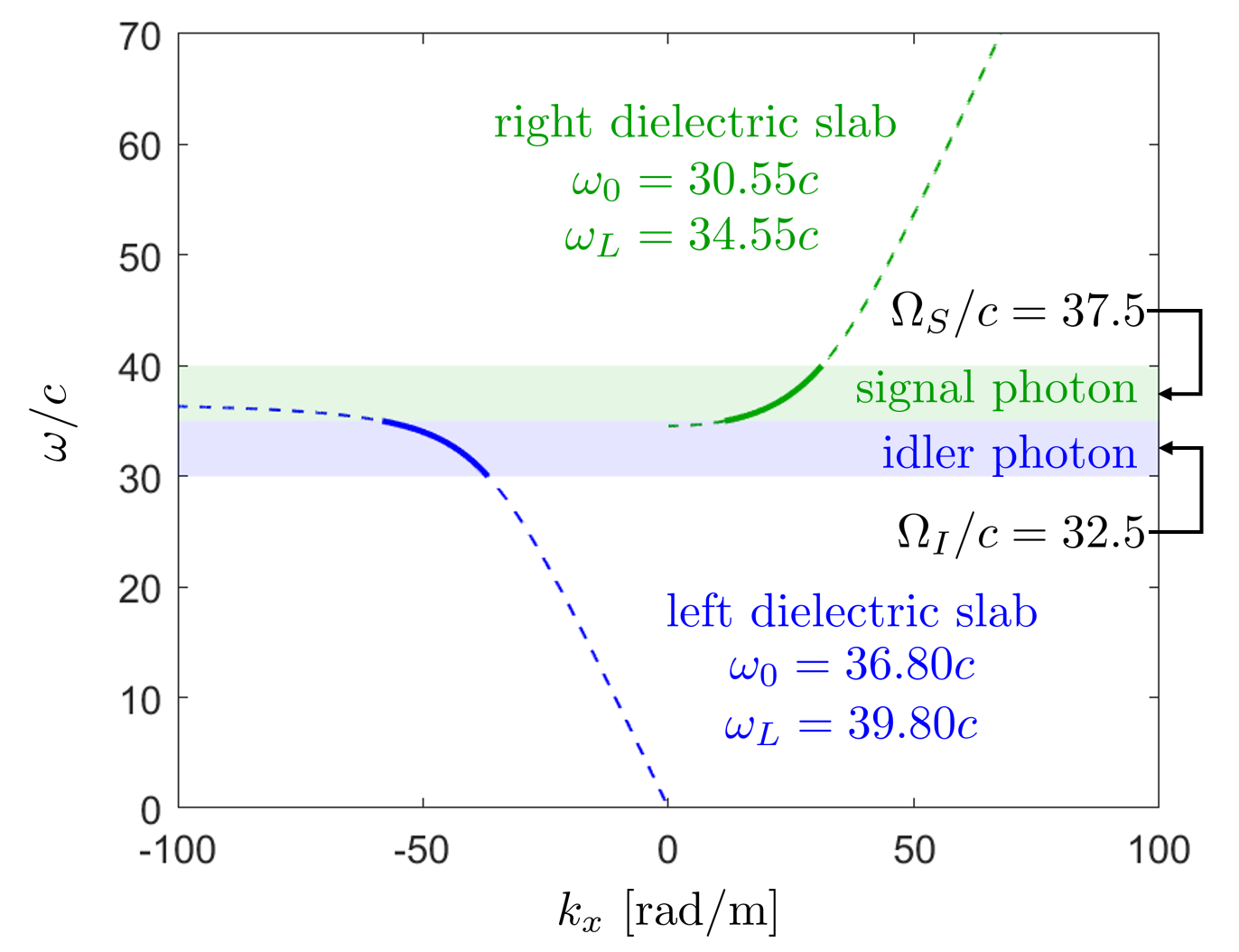}}
\\
\subfloat[second-order dispersion]{\includegraphics[width=1\linewidth]{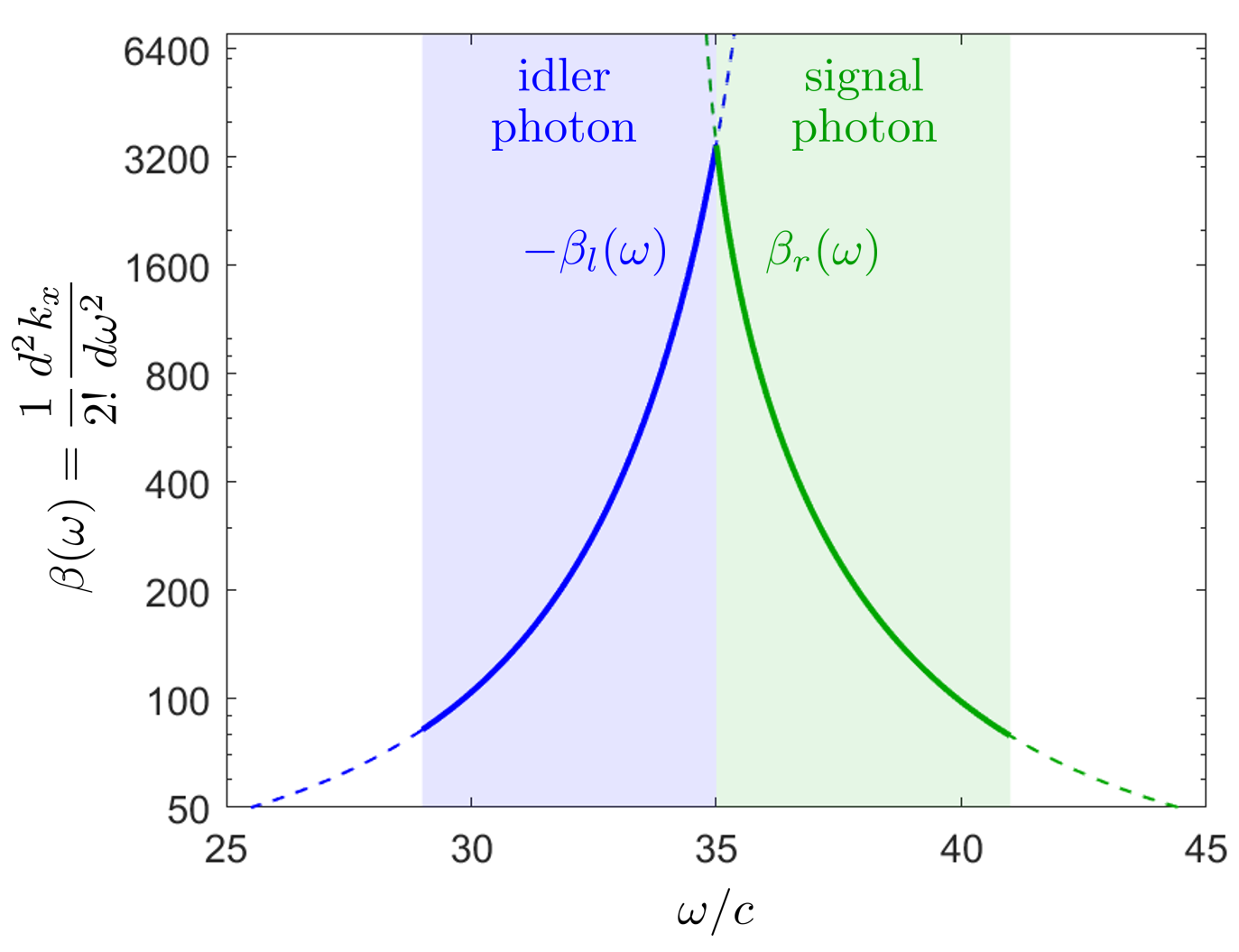}}
\caption{
Dispersion diagram of designed dispersive media. (a) dispersion diagram and (b) second-order dispersion $\beta$. The signal and idler photons will experience the same magnitude of the second-order dispersions with opposite sign.
}
\label{fig:dispersive_medium_modeling}
\end{figure}

To induce non-local dispersion cancellation, we introduce a dispersive medium composed of uniformly-filled single species of Lorentz oscillators.
And we exploit highly dispersive two local regimes in the dispersion diagram: One is below the bandgap for idler photons the other is above the bandgap for the signal photons, as illustrated in Fig. \ref{fig:dispersive_medium_modeling}(a).
By properly choosing resonant and plasma frequencies of the media, we can achieve the same magnitude of the second-order dispersion ($\beta$) with opposite signs over the photon's bandwidth, as illustrated in Fig. \ref{fig:dispersive_medium_modeling}(b).
The length of the both dispersive media is $L_{s}$.

\subsection{Coincidence}
\begin{figure}
\centering
\subfloat[$\psi(\omega_{2},\omega_{1})$]{\includegraphics[width=.5\linewidth]{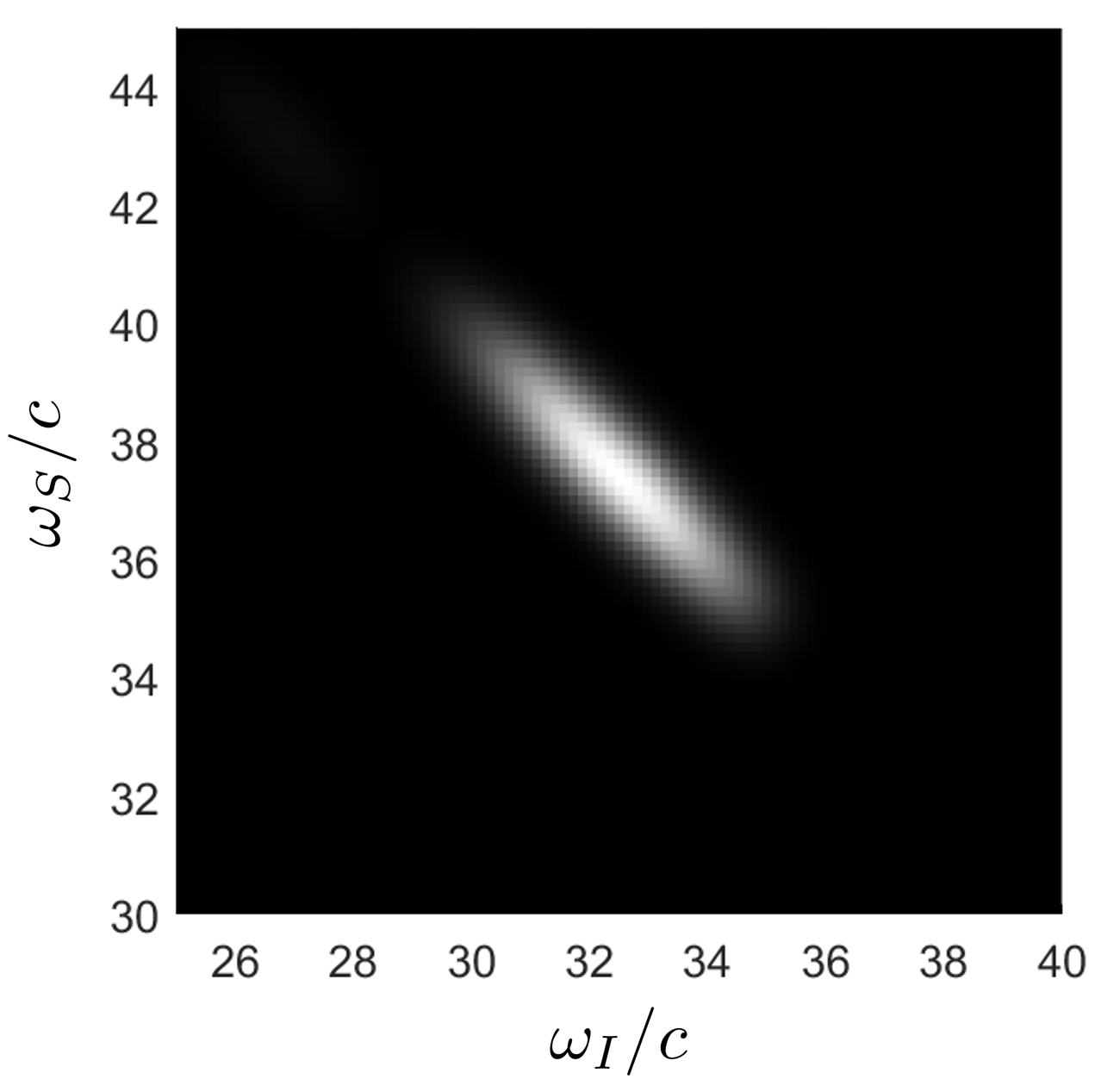}}
\subfloat[$\phi(\omega_{2},\omega_{1})$]{\includegraphics[width=.5\linewidth]
{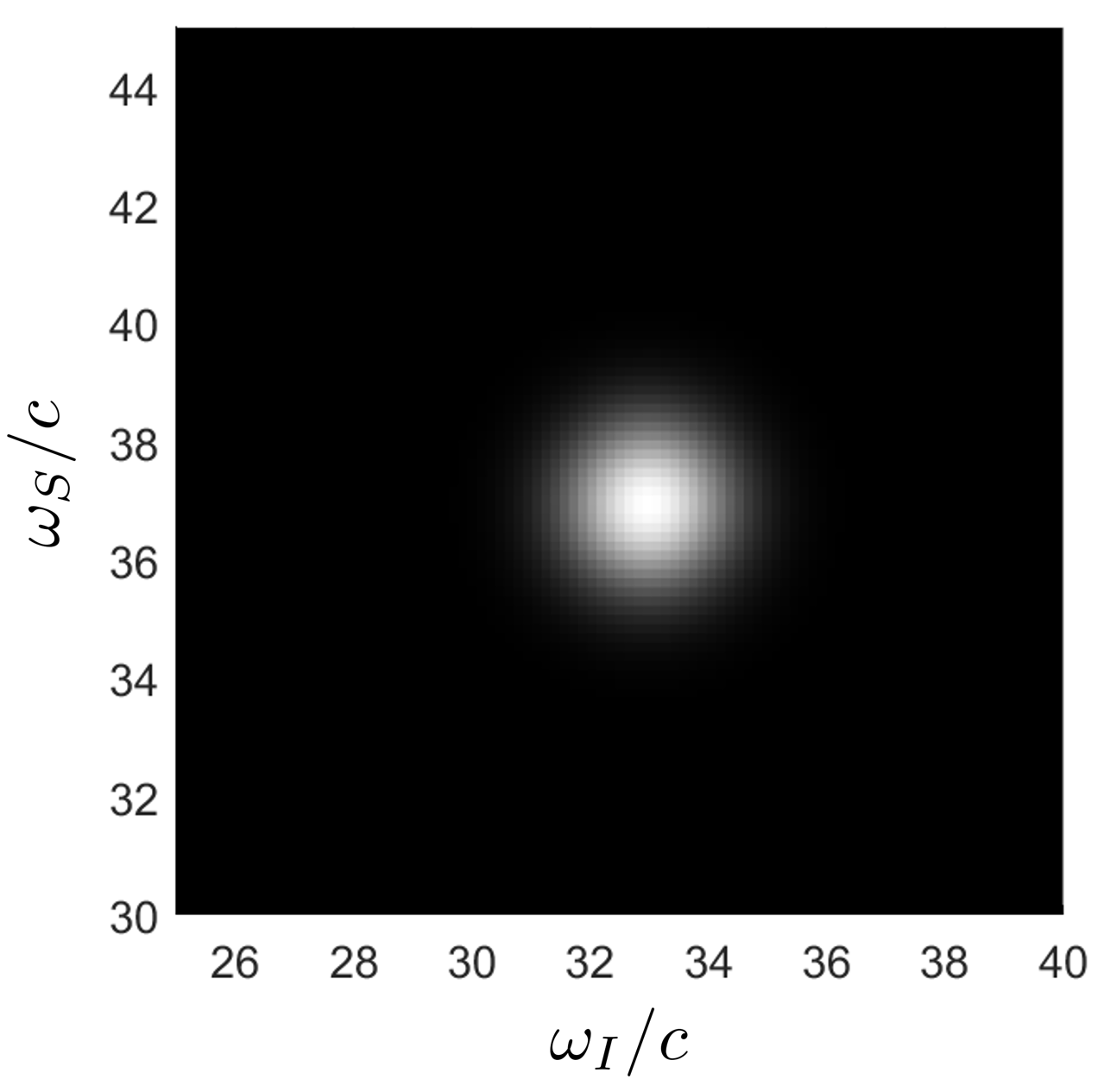}}
\\
\subfloat[two-time coincidence for entangled photon pair]{\includegraphics[width=.5\linewidth]{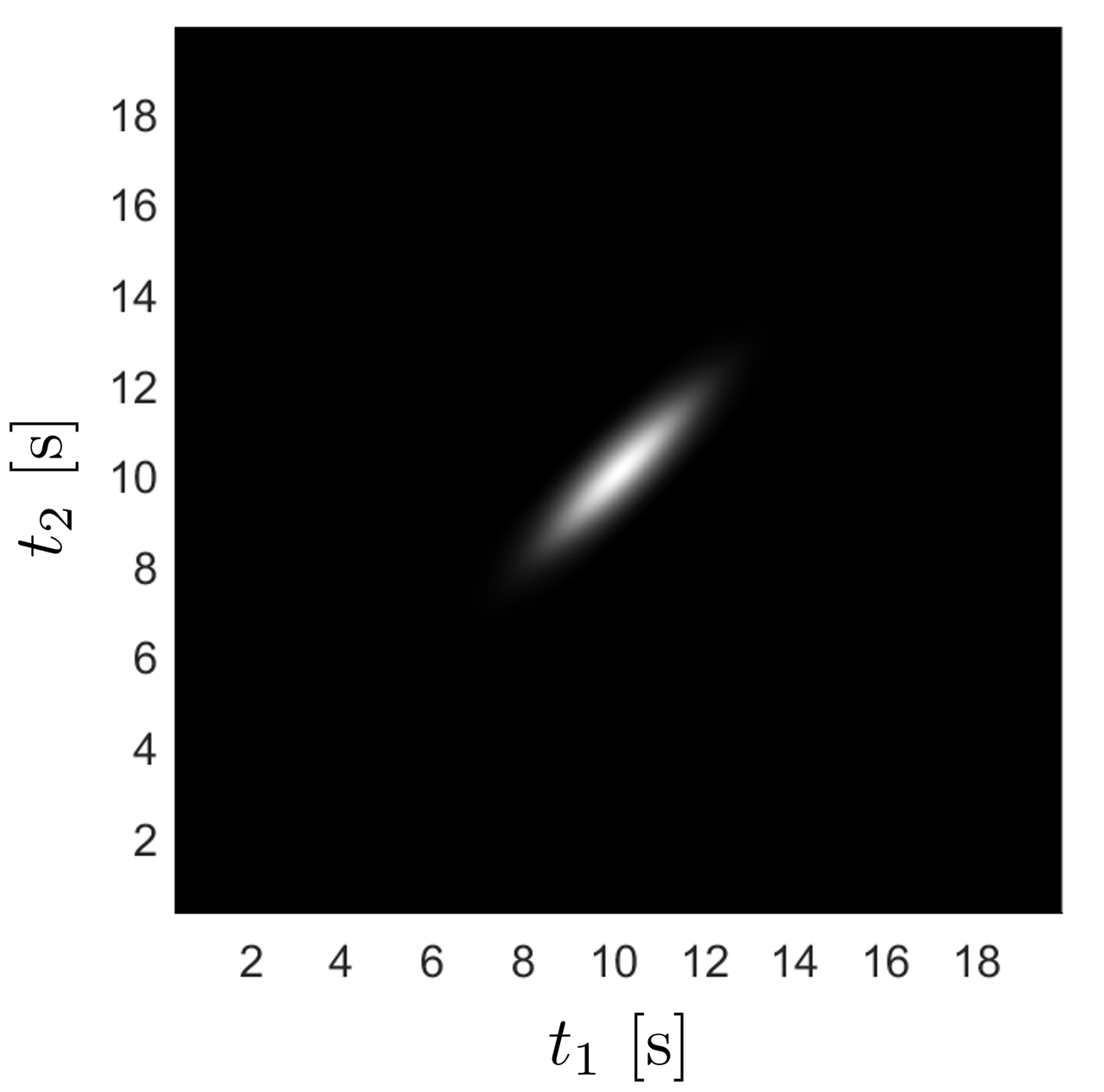}}
\subfloat[two-time coincidence for non-entangled photon pair]{\includegraphics[width=.5\linewidth]
{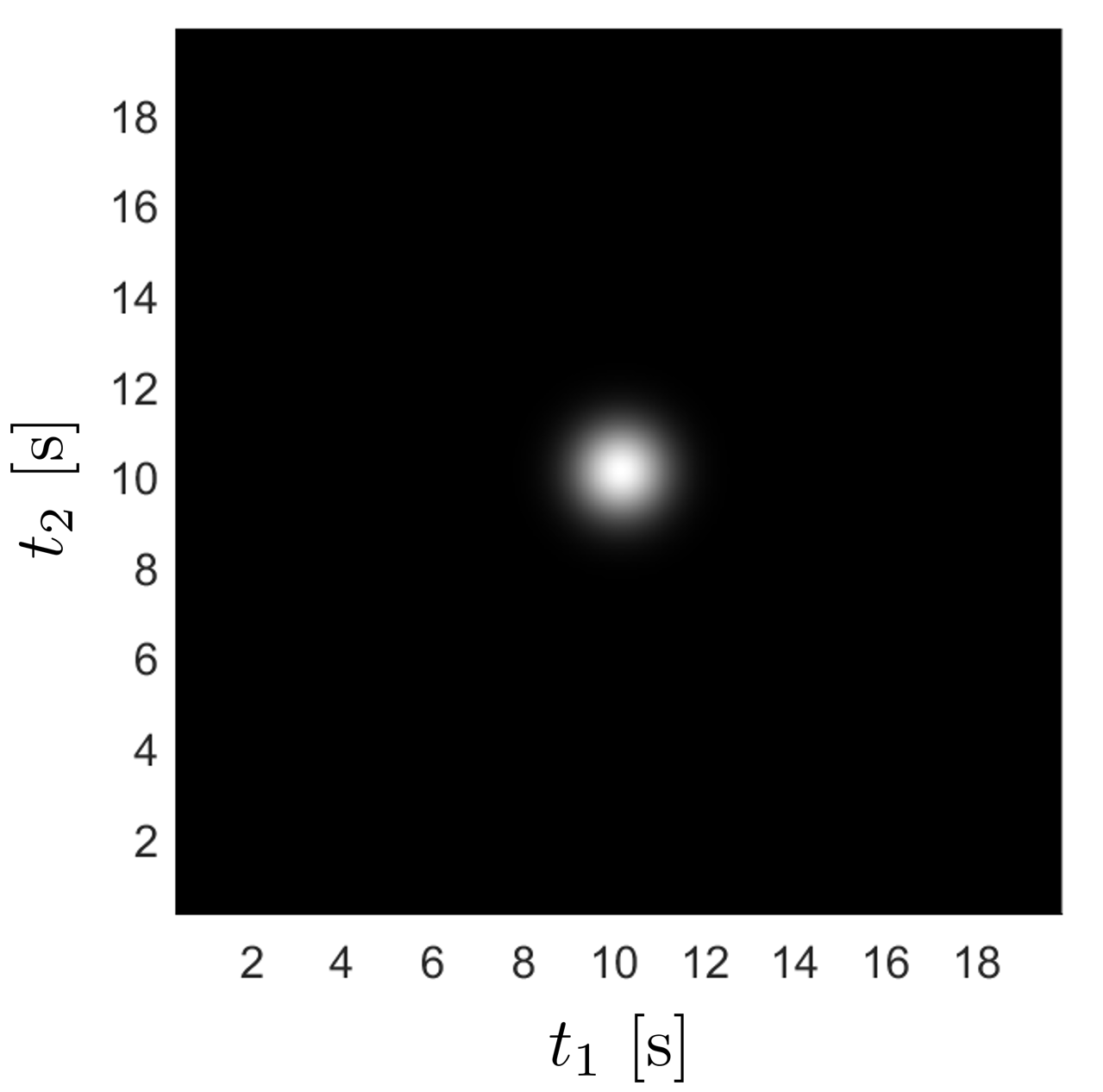}}
\caption{
Spectral probability amplitudes for (a) entangled two photons and (b) non-entangled two photons. The degree of coincidence for (c) entangled two photons and (d) non-entangled two photons. The entangled photon pair has both strong temporal correlation and frequency anticorrelation whereas non-entangled photon pair does not.
}
\label{fig:spectrum_coincidence_ETET_NET}
\end{figure}

We compute the degree of coincidence for the above two cases at different times $t_{1}$ and $t_{2}$, as depicted in Fig. \ref{fig:spectrum_coincidence_ETET_NET}(c) and Fig. \ref{fig:spectrum_coincidence_ETET_NET}(d), respectively.
One can observe that the entangled photon pair has both strong temporal correlation and frequency anticorrelation obeying
\begin{flalign}
\Delta\left(t_{2}-t_{1}\right)\Delta\left(\omega_{2}+\omega_{1}-\Omega_{P}\right)\leq 1
\end{flalign}
where $t_{2}$ and $t_{1}$ are detection times of the signal and idler photons, respectively.
However, the non-entangled photon pair does not exhibit any (anti)correlations.

\section{Consistency with past works}
We show that the present method is consistent with some of previous works in some limiting cases.

\subsection{Free field quantization}
In the vacuum, by using the Coulomb gauge with $\Phi=0$, there are non-zero dynamical variables $\mathbf{A}$ and $\boldsymbol{\Pi}_{AP}$.
Taking the no cross coupling description, one can easily check that \eqref{eqn:GHEVP_no_cross_coupling} is equivalent to the conventional Helmholtz wave equation for $\mathbf{A}$
\begin{flalign}
\nabla^{2}
\tilde{\mathbf{A}}_{\omega,\lambda}(\mathbf{r})
+
\omega^{2}\epsilon_{0}\mu_{0}\tilde{\mathbf{A}}_{\omega,\lambda}(\mathbf{r})
=0
\end{flalign}
where analytic solutions are plane waves.
The subsequent quantization can be easily done with the plane wave basis.
Note that plane waves has a dispersion relation $\omega^{2}\epsilon_{0}\mu_{0}=k^{2}=\left|\mathbf{k}\right|^{2}$ exhibiting the one to one correspondence between $\omega$ and $\mathbf{k}$.
Thus, a photon has both definite energy and momentum.

\subsection{Inhomogeneous dispersionless dielectric medium}\label{sec:Inhomo_Dispersionless}
The present methods are still valid when the background vacuum is replaced by dispersionless inhomogeneous medium, modeled by $\epsilon_{\infty}(\mathbf{r})$ and $\mu_{\infty}(\mathbf{r})$, making the resulting EVP still Hermitian.
Although this assumption does not satisfy the Kramers-Kronig relation over all frequencies, it
will be computationally efficient when dealing with a medium which is almost dispersionless over a narrow bandwidth.

Assume that Lorentz oscillators are absent (viz. no dispersive medium) and the background is filled by dispersionless and inhomogeneous dielectric medium.
Using the generalized Coulomb gauge with $\Phi=0$ and again taking the no coupling approach, one can show that \eqref{eqn:GHEVP_no_cross_coupling} is equivalent to the conventional Helmholtz wave for dispersionless inhomogeneous dielectric media; viz.,
\begin{flalign}
\nabla^{2}\tilde{\mathbf{A}}_{\omega,\lambda}(\mathbf{r})
+
\omega^{2}\epsilon_{\infty}(\mathbf{r})\mu_{0}\tilde{\mathbf{A}}_{\omega,\lambda}(\mathbf{r})
=
0.
\end{flalign}
In this case, time-harmonic eigenmodes, which corresponds to Bloch-Floquet modes, do not hold the one-to-one correspondence between $\omega$ and $\mathbf{k}$; hence, monochromatic photons cannot have a definite momentum.
Canonical quantization can be done with the Bloch-Floquet modes \cite{Glauber1991quantum,Knoll1987action,Na2020quantum}.

\subsection{1-D homogeneous dispersive medium}
A quantized vector potential field operator in 1-D vacuum uniformly filled by lossless Lorentz oscillators, which models a dispersive and homogeneous dielectric medium, was represented by \cite{Huttner_1991,Blow1990continuum}
\begin{flalign}
\hat{A}(x,t)
&=
\int_{\Omega_{+}}d\omega
\sum_{\lambda=-}^{+}
\Bigl(
\mathcal{A}_{0}e^{ik_{x}x}
\hat{c}_{\omega,\lambda}e^{-i\omega t}
+
\text{h.c.}
\Bigr),
\label{eqn:analytic_A_disp}
\\
\hat{\Pi}_{AP}(x,t)
&=
\int_{\Omega_{+}}d\omega
\sum_{\lambda=-}^{+}
\Bigl(
\mathcal{D}_{0}e^{ik_{x}x}
\hat{c}_{\omega,\lambda}e^{-i\omega t}
+
\text{h.c.}
\Bigr),
\end{flalign}
where $\lambda=\pm$ denotes the propagation direction degeneracy and
\begin{flalign}
\mathcal{A}_{0}
&=
-i\sqrt{\frac{\hbar v_{g}(\omega)}{4\pi \epsilon_{0}c \sqrt{\epsilon(\omega)}\omega}},\\
\mathcal{D}_{0}
&=
-\sqrt{\frac{\hbar v_{g}(\omega) \sqrt{\epsilon(\omega)}^{3}\omega}{4\pi\epsilon_{0}c}}
\end{flalign}
where $v_{g}=d\omega/d k_{x}$ denotes group velocity.
The dispersion relation is given by $k_{x}^{2}=\omega^{2}\epsilon(\omega)\mu_{0}$ where $\epsilon(\omega)=\left(1+\frac{\omega_p^2}{\omega_0^2-\omega^{2}}\right)\epsilon_{0}$; hence, it also destroys the one-to-one correspondence between $\omega$ and $k_{x}$.
The time-harmonic eigenmodes for $\hat{A}(x,t)$ and $\hat{\Pi}_{AP}(x,t)$ take the form of
\begin{flalign}
\tilde{A}_{\omega,\lambda}(x)=\mathcal{A}_{0}e^{i k_{x} x},\\
\tilde{\Pi}_{AP,\omega,\lambda}(x)=\mathcal{D}_{0}e^{i k_{x} x},
\end{flalign}
and their ratio becomes
\begin{flalign}
\mathcal{C}_{0}
=
\frac{\tilde{A}_{\omega,\lambda}(x)}{\tilde{\Pi}_{AP,\omega,\lambda}(x)}
=
\frac{\mathcal{A}_{0}}{\mathcal{D}_{0}}=\frac{i}{\omega\epsilon(\omega)}.
\end{flalign}

\begin{figure}
\centering
\includegraphics[width=\linewidth]
{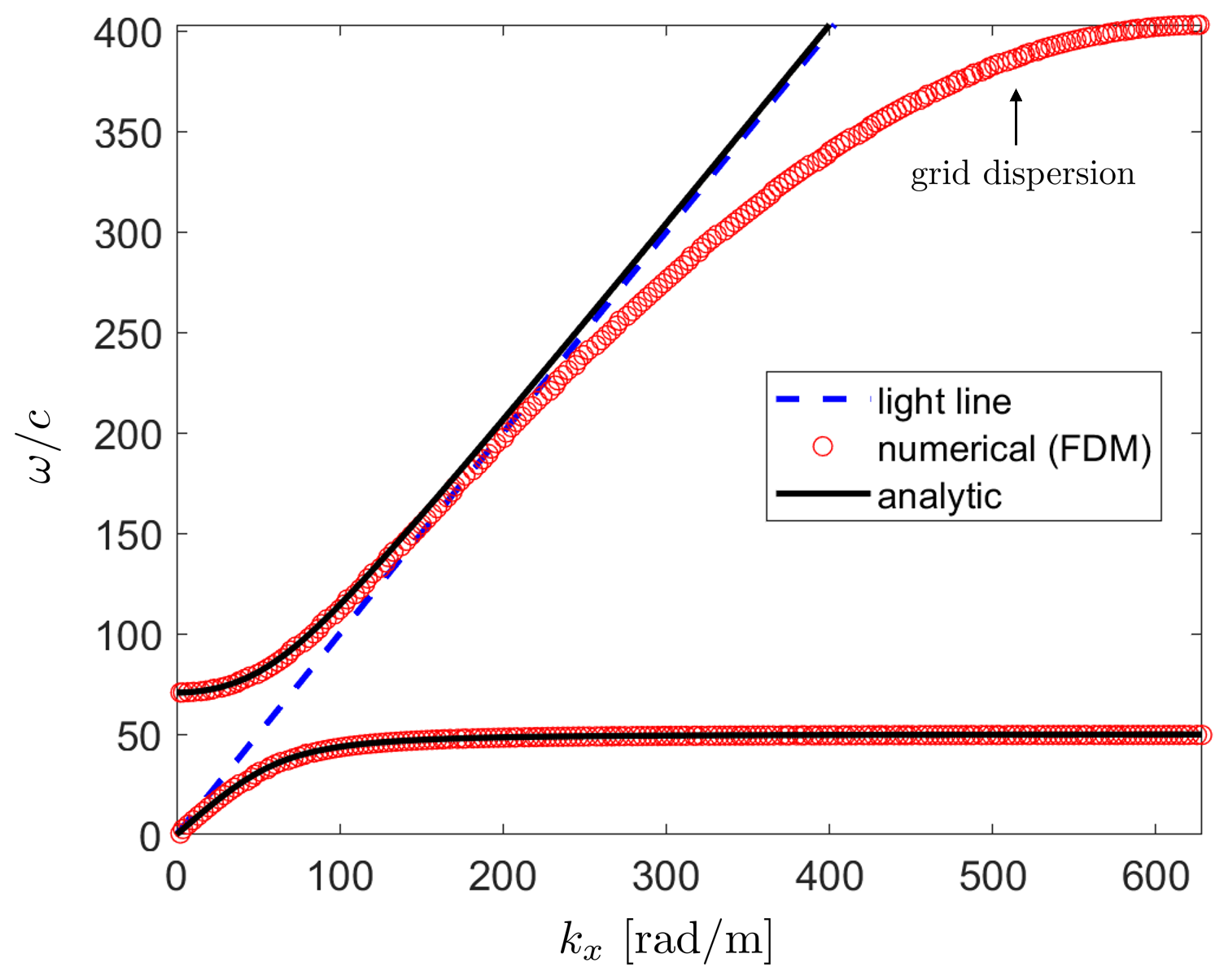}
\caption{
Dispersion relation $(\omega,k_x)$ for a dispersive and homogeneous dielectric medium ($\omega_{0}=\omega_{p}=50c)$.
Analytic (blue dashed line) and numerical (red circle) results are compared, showing great agreement except for in the high frequency regime.
This is due to the numerical dispersion effect which makes the phase and group velocities gradually slower \cite{taflove2000computational}.
}
\label{fig:HB_dispersion}
\end{figure}

\begin{figure}
\centering
\includegraphics[width=\linewidth]
{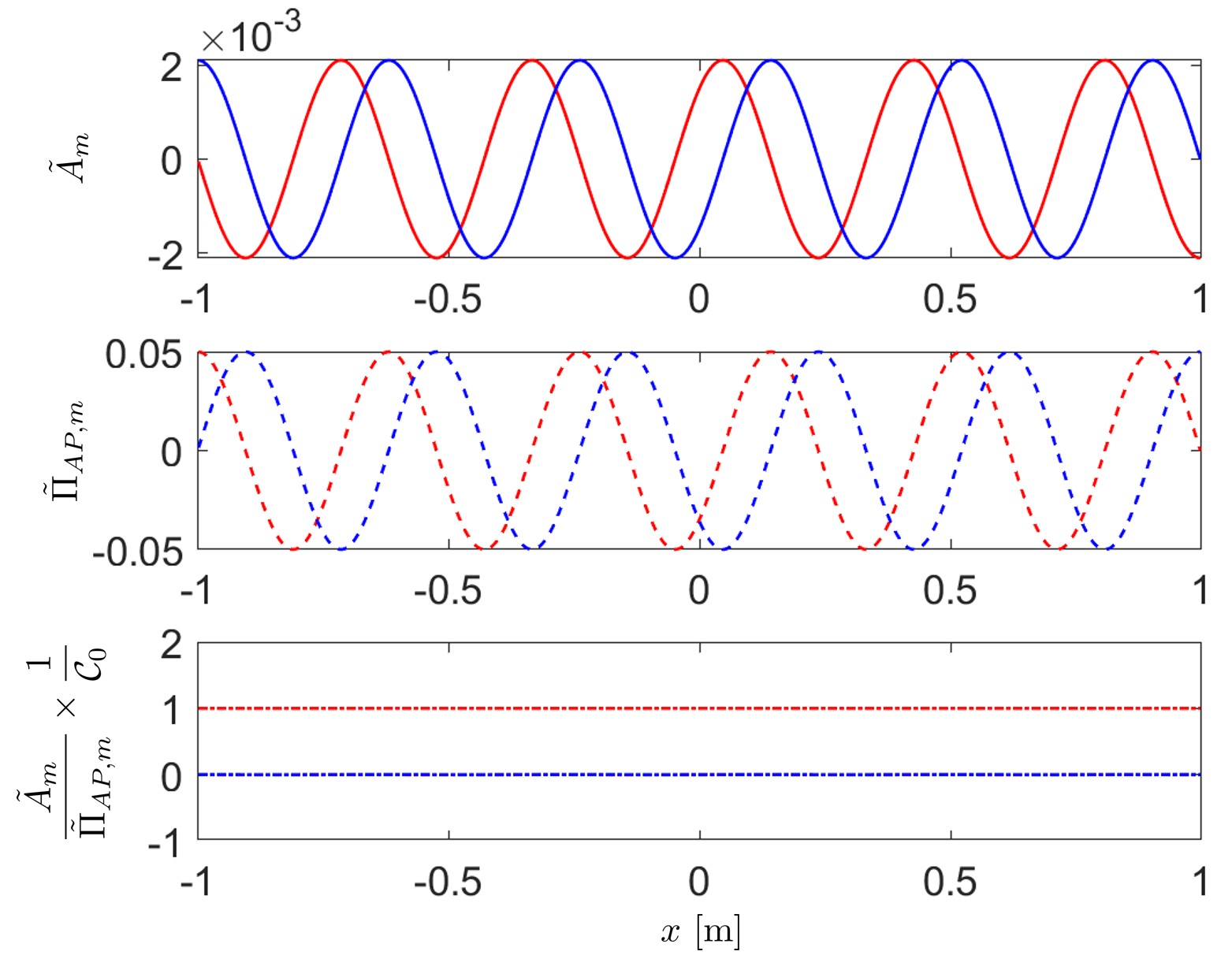}
\caption{
Illustration of $n=11^{\text{th}}$ numerical time-harmonic eigenmode for (a) vector potential and (b) its conjugate variable where $\omega_{n}\approx 11.52c$.
Note that red and blue curves represent real and imaginary values.
The ratio $\tilde{A}_{m}/\tilde{\Pi}_{AP,m}$ normalized by $\mathcal{C}_0=i/(\omega\epsilon(\omega_m))$ is displayed in (c), showing the real and imaginary values are unity and zero, respectively.
}
\label{fig:numerical_temporal_eigenmode}
\end{figure}

Taking the cross coupling description, we use the finite-difference method (FDM), which is widely used across all scientific areas due to simplicity and reliability, to obtain a set of numerical time-harmonic eigenmodes.
The dispersive medium was modeled by Lorentz oscillators (having $\omega_{p}=\omega_{0}=50c$) uniformly filled over the entire problem domain $x\in\left[-L/2,L/2\right]$.
Note that the domain was discretized by $201$ number of grid points.
Since the present method does not specify wavenumber of time-harmonic eigenmodes, we performed the spatial fast Fourier transform analysis to extract $k_{x}$ for each numerical time-harmonic eigenmode.
The dispersion relations of analytic and numerical time-harmonic eigenmodes are compared in Fig. \ref{fig:HB_dispersion}.
There is great agreement between them except for in the high frequency regime.
The deviation comes from the numerical grid dispersion error in using the finite-difference approximation \cite{taflove2000computational}.
This can be mitigated by using advanced CEM methods, such as finite-element or pseudo-spectral methods.
It is observed that, at a given $k_{x}$, there are two plane wave solutions having different eigenfrequencies lying on lower and upper branches, respectively.
The gap between lower and upper branches, i.e., $\omega\in\bigl(\omega_0,\sqrt{\omega_0^2+\omega_p^2}\bigr)$ is related to the anomalous dispersion region if absorption is included \cite{Huttner_1991}.
Fig. \ref{fig:numerical_temporal_eigenmode} illustrates the $n=11$-th numerical time-harmonic eigenmodes.
One can clearly observe that the ratio between $\tilde{A}_{m}$ and $\tilde{\Pi}_{AP,m}$ maintains $\mathcal{C}_{0}$, which is consistent with the analytic time-harmonic eigenmodes.

\subsection{Purcell factor in dispersive medium}
\begin{figure}
\centering
\includegraphics[width=\linewidth]
{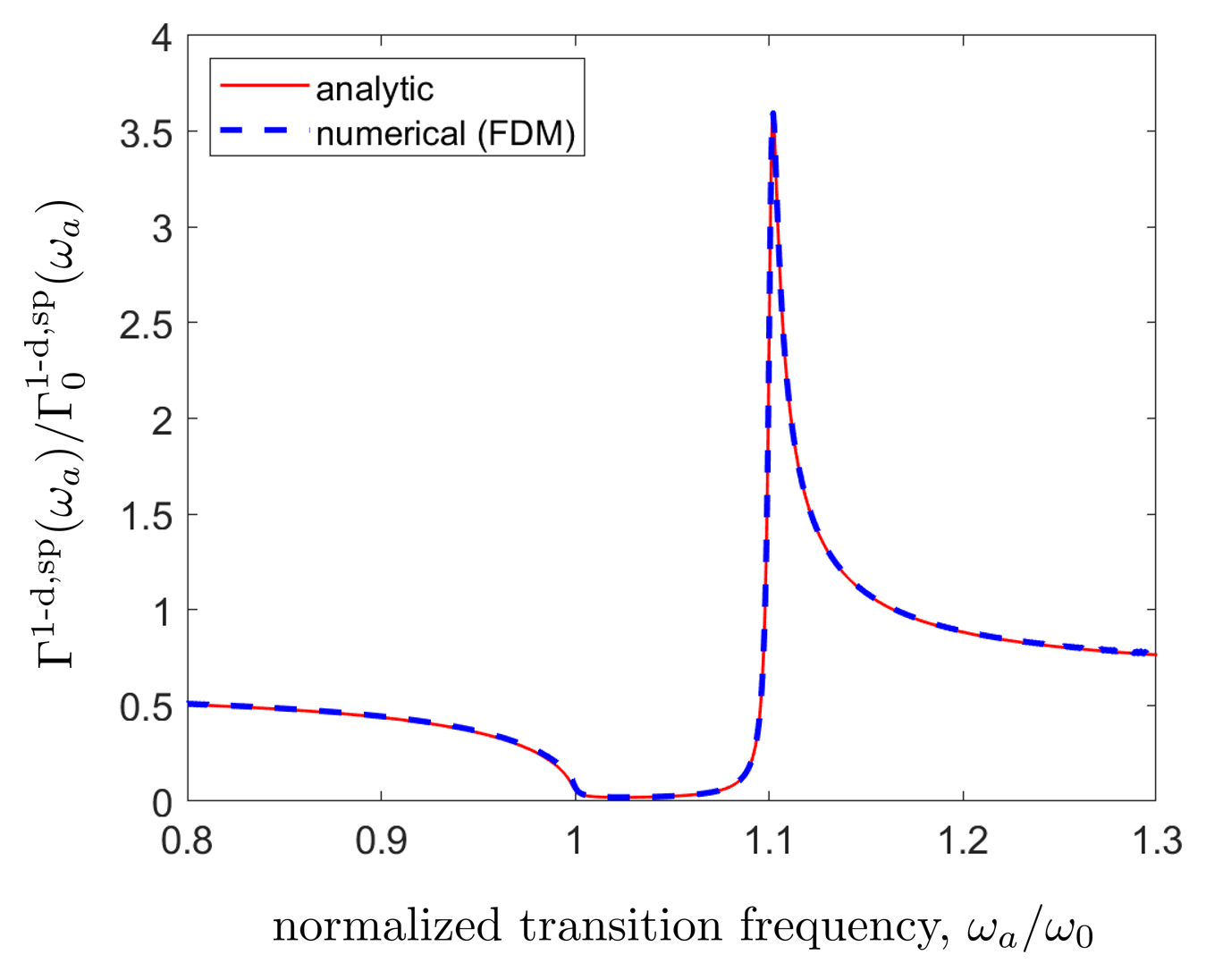}
\caption{
Purcell factors versus normalized transition frequency $\omega_a$ when a two-level system is embedded in 1-D dispersive medium with a single polarization.
The SER can be enhanced above the bandgap where $
\omega\in\left[\omega_0,~ \sqrt{\omega_0^2+\omega_p^2} \right]$.
}
\label{fig:SER_1D_dispersive}
\end{figure}
Spontaneous emission rate (SER) of an excited atom can be enhanced by introducing cavity \cite{PhysRevLett.50.1903}, plasmonic structures \cite{Chen2012study}, photonic crystals \cite{Englund2005controlling}, and dielectric media \cite{Barnett1992spontaneous}.
To check the validity of the proposed method, we compute a Purcell factor when an excited two-level system is embedded in a dispersive medium.
For simplicity, we consider the 1-D free space along $x$-axis and single polarization (electric field operators are polarized along $y$-axis).
The SER of this system can be determined by the Fermi golden rule \cite{fox2006quantum}
\begin{flalign}
\Gamma^{\text{1-d,sp}}(\omega_{a})
=
\frac{2 \mu_{a}^{2}}{3 c \hbar^{2}}
\mel
{0}
{\hat{E}^{(+)}_{a}
\hat{E}^{(-)}_{a}
}
{0}
\delta(\omega-\omega_{a})
\end{flalign}
where $\omega_{a}$ and $\mu_{a}$ denote transition frequency and dipole moment, and
\begin{flalign}
\hat{E}^{(+)}_{a}
=
\hat{E}^{(+)}(x_{a},t)
=
i\omega\hat{A}^{(+)}(x_{a},t)
\end{flalign}
at the two-level system's location $x_{a}$.
We analytically calculate the above SER by using \eqref{eqn:analytic_A_disp} and approximating $\delta(\omega-\omega_{a})$ by Lorentzian distribution.
In addition, we evaluate the SER by using numerical time-harmonic eigenmodes based on the no cross coupling description.
Fig. \ref{fig:SER_1D_dispersive} compares Purcell factors versus normalized transition frequency for analytic and numerical time-harmonic eigenmodes.
Note that the SER for the 1-D free-space with a single polarization is calculated from the vacuum fields and density of states (DOS) \cite{fox2006quantum}
\begin{flalign}
\Gamma_{0}^{\text{1-d,sp}}(\omega_{a})
=
\frac{\mu_{a}^{2}\omega_{a}}{3\hbar\epsilon_{0}c}.
\end{flalign}
There is great agreement between two results, which successfully validates the proposed method.
The SER can be enhanced above the bandgap where $
\omega\in\left[\omega_0,~ \sqrt{\omega_0^2+\omega_p^2} \right]$.

\subsection{1-D dispersive dielectric slab surrounded by the free space}
\begin{figure}
\centering
\includegraphics[width=\linewidth]
{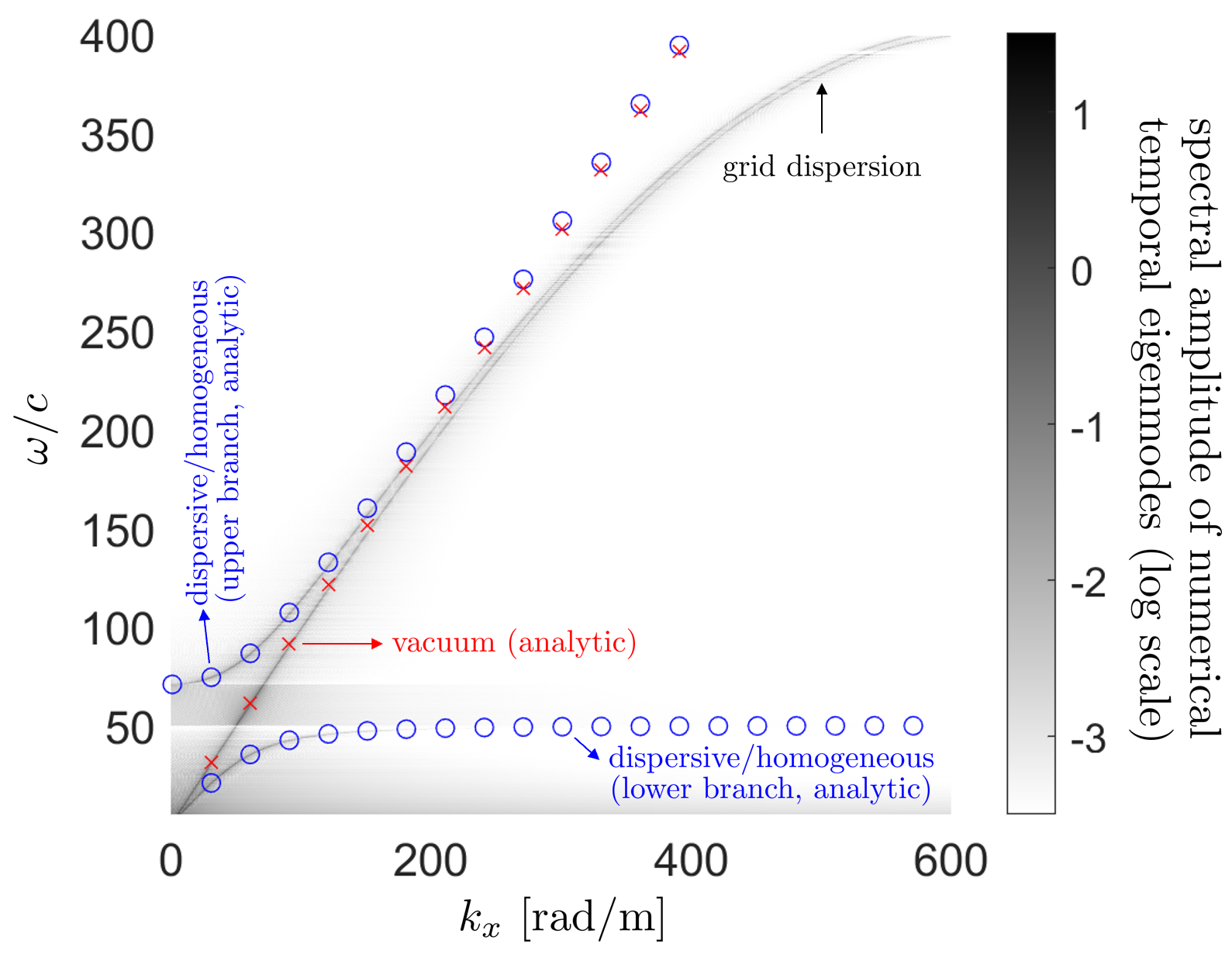}
\caption{
Dispersion relation $(\omega,k_x)$ for the dispersive dielectric slab.
The contour map (gray color scale) illustrates spectral amplitudes of numerical time-harmonic eigenmodes.
There are three plane wave solutions having different eigenfrequencies at a given $k_{x}$.
Two sets of dispersion relations can be observed: One is of the vacuum (red X marker) and the other is of the dispersive and homogeneous dielectric medium (blue O marker).
}
\label{fig:intermediate}
\end{figure}

Finally, we consider a dispersive dielectric slab (sized by $x\in\left[-L/4,L/4\right]$) inside a vacuum box sized by $x\in\left[-L/2,L/2\right]$.
We take the cross coupling description to obtain numerical time-harmonic eigenmodes and compute their spectral amplitudes by performing the spatial fast Fourier transform analysis.
The result is displayed in Fig. \ref{fig:intermediate}, compared with dispersion relations of the vacuum and dispersive and homogeneous dielectric medium.
There are three plane wave solutions with different eigenfrequencies at a given $k_{x}$.
Two sets of dispersion relations can be observed: One is of the vacuum (red X marker) and the other is of the dispersive and homogeneous dielectric medium (blue O marker).
Hence, it is between two limiting cases: (1) vacuum when the slab width converges to $0$ and (2) dispersive and homogeneous medium when the whole vacuum box is filled by the slab, as expected.

\bibliography{mybibpra}

\end{document}